%% file: paper.tex
\documentclass[fleqn,usenatbib]{mnras}

\usepackage{newtxtext,newtxmath}

\usepackage[T1]{fontenc}
\usepackage{ae,aecompl}
\usepackage{caption}
\usepackage{graphicx}	
\usepackage{amsmath}	
\usepackage{amssymb}	
\usepackage{subcaption}

\newcommand{\plya}{\,$\mathrm{P_{Ly\alpha}}$}
\newcommand{\lya}{\,$\mathrm{Ly\alpha}$}
\newcommand{\msol}{\, $\mathrm{M } _ { \odot }$}
\newcommand{\ragn}{\,$\mathrm {r} _ {\mathrm{AGN}}$}
\newcommand{\ef}{\,$\mathrm{\epsilon_{\mathrm{f}}}$}
\newcommand{\hagn}{\,$\mathrm{HAGN}$}
\newcommand{\hnagn}{\,$\mathrm{HnoAGN}$}
\newcommand{\hlmin}{\,$\mathrm{HAGN-lmin11}$}
\newcommand{\hnlmin}{\,$\mathrm{HnoAGN-lmin11}$}
\newcommand{\clp}{\,$\mathrm{HAGN\mathrm{clp10}}$}
\newcommand{\clpp}{\,$\mathrm{HAGN\mathrm{clp100}}$}
\newcommand{\rp}{\,$\mathrm{HAGN\mathrm {r+}}$}
\newcommand{\hrm}{\,$\mathrm{HAGN\mathrm {r-}}$}
\newcommand{\ep}{\,$\mathrm{HAGN\mathrm {\epsilon+}}$}
\newcommand{\hem}{\,$\mathrm{HAGN\mathrm {\epsilon-}}$}

\newcommand{\ed}{\,$\mathrm{10^{-2}}$}
\newcommand{\et}{\,$\mathrm{10^{-3}}$}
\newcommand{\edd}{\,$\mathrm{10^{2}}$}

\title[AGN feedback on the P1D]{The impact of AGN feedback on the 1D power spectra from the Ly$\alpha$ forest using the Horizon-AGN suite of simulations}

\author[S. Chabanier et al.]{Sol\`ene Chabanier$^{1,2}$\thanks{E-mail: solene.chabanier@cea.fr},
Fr\'ed\'eric Bournaud$^{1,2}$,
Yohan Dubois$^{3}$,
Nathalie Palanque-Delabrouille$^{1}$,
\newauthor
Christophe Y\`eche$^{1}$,
Eric Armengaud$^{1}$,
S\'ebastien Peirani$^{3,4}$,
Ricarda Beckmann$^{3}$
\\
$^{1}$IRFU, CEA, Universit\'e Paris-Saclay, F91191 Gif-sur-Yvette, France\\
$^{2}$ AIM, CEA, CNRS, Universit\'e Paris-Saclay, Universit\'e Paris Diderot, Sorbonne Paris Cit\'e, F91191 Gif-sur-Yvette, France\\
$^{3}$CNRS and UMPC Universit\'e Paris 06, UMR 7095, Institut d'Astrophysique de Paris, 98 bis boulevard Arago, Paris 75014, France\\
$^{4}$ Universit\'e C\^ote d'Azur, Observatoire de la C\^ote d'Azur, CNRS, Laboratoire Lagrange, Nice, France\\
}

\date{Accepted XXX. Received YYY; in original form ZZZ}

\pubyear{2020}

\begin{document}
\label{firstpage}
\pagerange{\pageref{firstpage}--\pageref{lastpage}}
\maketitle

\begin{abstract}
The Lyman-$\alpha$ forest is a powerful probe for cosmology, but it is also strongly impacted by galaxy evolution and baryonic processes such as Active Galactic Nuclei (AGN) feedback, which can redistribute mass and  energy on large scales. We constrain the signatures of AGN feedback on the 1D power spectrum of the Lyman-$\alpha$ forest using a series of eight hydro-cosmological simulations performed with the Adaptative Mesh Refinement code RAMSES. This series starts from the Horizon-AGN simulation and varies the sub-grid parameters for AGN feeding, feedback and stochasticity. These simulations cover the whole plausible range of feedback and feeding parameters according to the resulting galaxy properties. AGNs globally suppress the Lyman-$\alpha$ power at all scales. On large scales, the energy injection and ionization dominate over the supply of gas mass from AGN-driven galactic winds, thus suppressing power. On small scales, faster cooling of denser gas mitigates the suppression. This effect increases with decreasing redshift. We provide lower and upper limits of this signature at nine redshifts between $z=4.25$ and $z=2.0$, making it possible to account for it at post-processing stage in future work given that running simulations without AGN feedback can save considerable amounts of computing resources. Ignoring AGN feedback in cosmological inference analyses leads to strong biases with 2\% shift on $\sigma_8$ and 1\% shift on $n_s$, which represents twice the standards deviation of the current constraints on $n_s$.
\end{abstract}

\begin{keywords}
(galaxies:) quasars: supermassive black holes -- (galaxies:)intergalactic medium -- (galaxies:) quasars: absorption lines
\end{keywords}



\section{Introduction}
\label{sec:intro}
\input{intro}

\section{The simulations set}
\label{sec:simu}
\input{simu}

\section{The 1D flux power spectrum}
\label{sec:p1d}
\input{p1d}

\section{Results}
\label{sec:results}
\input{results}

\section{Discussion}
\label{sec:discussion}
\input{discussion}

\section{Acknowledgements}
This work has been carried out thanks to the support of the ANR 3DGasFlows (ANR-17-CE31-0017). This work was granted access to the HPC resources of CINES and TGCC under the allocations 2019-A0070402192 and  2019-A0070410560  made by GENCI.




\bibliographystyle{mnras}
\bibliography{paper.bib} 


\bsp	
\label{lastpage}
\end{document}

%% file: intro.tex
Neutral hydrogen in the intergalactic medium (IGM) scatters light at 1216 $\mathring{A}$, producing characteristic absorption features in the spectra of distant quasars (QSO), dubbed the Lyman-$\mathrm{\alpha}$ (Ly$\mathrm{\alpha}$) forest. It becomes an increasingly used cosmological probe as it traces density fluctuations, ionization state and temperature of the IGM on a unique range of redshifts and scales, from few Mpc to hundreds of Mpc. On the one hand, 3D correlations in the Ly$\mathrm{\alpha}$ transmission field \citep{Slosar2011,Slosar2013,Bautista2017,duMasdesBourboux2017,deSainteAgathe2019}, that use informations from different lines-of-sight, but also cross-correlation with other tracers \citep{Perez2017,Blomqvist2019} accurately measure the position of the Baryon Accoustic Peak and provide constraints on dark matter and dark energy.
On the other hand, informations of the smallest scales are accessible through correlations along the line of sight, or equivalently through the 1D power spectrum \citep{McDonald2006,PalanqueDelabrouille2013,Chabanier2019a}. The latter is sensitive to the thermal state of the IGM and can therefore set constraints on reionization \citep{Zaldarriaga2001,Meiksin2009,Lee2015,McQuinn2016} and on its thermal history \citep{Viel2006,Bolton2008}.
But it is also sensitive to the clustering properties of small scales structures, and is therefore a remarkable probe of the impact of the density perturbation smoothing  caused by the free-streaming of relativistic particles such as neutrinos. It has been used to set constraints on the sum of the neutrinos masses \citep{ PalanqueDelabrouille2015a,PalanqueDelabrouille2015b,Yeche2017,PDB2019}, on warm   \citep{Baur2016,Yeche2017,Baur2017} and fuzzy dark matter models \citep{Armengaud2017,Irsic2017}. Finally, thanks to the unique range it probes, it was used  as one of the multiple independent cosmological probes to estimate the 3D matter power spectrum at $z \simeq 0$ at the Mpc scale \citep{Chabanier2019b}.

Thanks to the advent of spectroscopic surveys, an increasing amount of Ly$\mathrm{\alpha}$ data becomes available, with e.g. the BOSS survey \citep{Dawson2013} of SDSS-III \citep{eisenstein2011}, the eBOSS survey \citep{Dawson2016} of SDSS-IV \citep{Blantonsdss}, Keck/HIRES \citep{Viel2008,Viel2013} or VLT/UVES \citep{Walther2017} spectra. Because of a large number of spectra and an improved census of the systematic uncertainties of the analysis, measurements of the 1D power spectrum of the Ly$\mathrm{\alpha}$ forest have reached accuracy at the percent level \citep{Chabanier2019a}. And it will even shrink further with the Dark Energy Spectroscopic Instrument (DESI) \citep{desi2016} that will observe more than two million high redshift QSOs, or the WEAVE-QSO survey \citep{Pieri2016}.

The significant increase in statistical power must be combined with an improvement in the theoretical predictions from simulations. If it was thought that gravitational instabilities and hydrodynamics alone were the only significant processes to accurately model the Ly$\alpha$ forest, we now know that astrophysical processes can impact the thermal state and distribution of gas in the IGM. Indeed IGM gas might be consequently affected by galaxy evolution because of gas cooling, star formation and the feedbacks from Supernovae (SN) and Active Galactic Nuclei (AGN) that inject energy in the ambient medium. In order to fully exploit the small-scale data, it is necessary to have these non-linear effects under control.

It has been shown that AGN feedback strongly impact properties and evolution of galaxies \citep{Silk98,Croton2006,Kaviraj2015,Dubois2014,Illustris2017} but also many cosmological observables such as the total matter power spectrum \citep{Chisari2018,Barreira2019}, the matter bispectrum \citep{Foreman2019}, the orientation of the spin of galaxies \citep{Dubois2014b}, the density profile of dark matter haloes and galaxies \citep{Peirani2008,Martizzi2013,Schaller2015,Peirani2017,Peirani2019}.

To explore the impacts of AGN processes numerous approaches have been proposed, recently summarized in \cite{Chisari2019}. Semi-analytic methods \citep{Croton2006,Bower2006} such as modifications of the halo model or baryonification methods have the advantage to be fast and can be used to explore the parameter space in inference analysis. However, they present the main drawback to lack accuracy, so they need to be calibrated against hydrodynamical simulations or observations. Also, major efforts have been conducted to implement AGN feedback in hydrodynamical simulations \citep{Sijacki2007,diMatteo2008,Dubois2012}, leading to massive cosmological simulations e.g. Horizon-AGN \citep{Dubois2014}, IllustrisTNG \citep{Illustris2017}, EAGLE \citep{Schaye2014}, the OWLS project \citep{Schaye2010} or MassiveBlack-II \citep{Khandai2015} in order to reproduce some key-galaxy properties.

In this paper we quantify the impact of AGN feedback on the 1D power spectrum from the Ly$\alpha$ forest and give analytical corrections making it possible to account for AGN effects at post-processing stage in future works. Previous studies showed that the Ly$\alpha$ forest gas should not be consequently impacted by feedback processes because winds tend to escape into voids \citep{Bertone2006,Tornatore2010,Tepper2012}. In \cite{Viel2012}, hereafter V13, a set of hydrodynamical simulations from the OWLS project \citep{Schaye2010} was used to estimate the impact of AGN feedback and other baryonic effects such as metal line cooling or stellar winds on the Ly$\mathrm{\alpha}$ flux statistics. The authors show that AGN feedback have significant effects  compared to the statistical uncertainties at that time. However, while the study explored several scenarios of SN-driven winds, the analysis is based on a specific AGN feedback model because they use one set of parameters for the AGN feedback, coupled with one specific hydrodynamical code  used at the OWLS simulation resolution. In this work we extend the analysis done in V13 by giving a correction on the 1D power spectrum of the Ly$\mathrm{\alpha}$ forest with uncertainties that encompass the whole range of plausible feedback models in a range of scales and redshifts that DESI will reach.
To do so, we use the Horizon-AGN (\hagn \,) simulation \citep{Dubois2014} with a more refined resolution in the diffuse IGM that consitutes the Ly$\alpha$ forest gas, which we describe in Sec.~\ref{sec:HAGN}. We construct a series of additional simulations, presented in Sec.~\ref{sec:variants} with a set of feedback and feeding parameters spanning the observational uncertainties of galaxy properties  at $z = 2$. In Sec.~\ref{sec:method} we outline the numerical methods used to derive the flux power spectra from the simulations and we present how well the Ly$\alpha$ forest is reproduced in the Horizon-AGN simulation in Sec.~\ref{sec:lya}. We estimate the impact of AGN feedback  on the 1D power spectrum in Sec.~\ref{sec:correction}, but we also put an upper and lower bound on the correction to span the whole range of plausible subgrid parameters in Sec.~\ref{sec:uncertainties}. Finally we estimate the impact of using these new corrections on estimation of cosmological parameters in Sec.~\ref{sec:cosmo_impact}.

%% file: simu.tex
In Sec.~\ref{sec:HAGN} we present the \hagn \, simulation chosen as the fiducial simulation, and in Sec.~\ref{sec:variants} we present the suite of simulations we will use in Sec.~\ref{sec:results} to estimate the uncertainties related to the feedback model in our corrections. We vary parameters in the subgrid model to cover a wide range of realistic models that span the observational uncertainties related to the BH-to-galaxie mass relation ($\mathrm{M_{BH}}-\mathrm{M_*}$) and to the mean fraction of gas in galaxies.

The choice of \hagn \, as our fiducial simulation was motivated by the following arguments. First, the grid-based method, i.e. Eulerian method, is necessary to control the resolution in the lowest density regions of the IGM that constitute the Ly$\alpha$ forest. Indeed this is not possible with smoothed particle hydrodynamics (SPH), i.e. Lagrangian methods, that spend time evolving the highest density regions. We also need a large enough box, not only not to miss the large-scale modes, but also because of the non-linear coupling of modes during gravitational evolution. \cite{Tytler2009} also state that small boxes are too cold compared to larger boxes beause of shock heating being not frequent enough in the small boxes. \cite{Lukic2014} and \cite{Borde2014} used boxes of 80 $ \mathrm{ Mpc.h^{-1}}$ to capture all scales when \cite{Bolton2009} and \cite{McDonald2003} used 40 $ \mathrm{ Mpc.h^{-1}}$. With a 100 $ \mathrm{ Mpc.h^{-1}}$ box, \hagn \, is a conservative choice.
For the resolution \cite{Lukic2014} requires a 20 $ \mathrm{ kpc.h^{-1}}$ cell size for a converged Ly$\alpha$ power spectrum on an uniform mesh without AMR. However, the implementation of baryonic physics with AGN and stellar feedbacks that accurately reproduces properties of galaxy evolution requires a resolution at the kpc scale. Such a dynamical range is computationally too demanding. Using a more refined simulation in the diffuse regions, we show in Sec.~\ref{sec:results} that the 100 $\mathrm{ kpc.h^{-1}}$ maximal cell size in \hagn \, appears to be sufficient to quantify the effects of AGN feedback because we need the numerical convergence of the correction and not the absolute value of the power spectra. We performed tests on the convergence of the correction that are presented in Sec.~\ref{sec:results}. Finally, we want to include uncertainties in the feedback model by varying the main subgrid parameters in order to have $\mathrm{M_{BH}}-\mathrm{M_*}$ and the mean fraction of gas in the range of the observational uncertainties. Therefore our fiducial simulation should be in agreement with observations. \hagn \, was calibrated at $z = 0$ on the Maggorian relation \citep{Dubois2012, Volonteri2016} and appears to reproduce the observed fractions of gas in galaxies at different redshifts, which is one of the main issue in other cosmological hydrodynamical simulations. On the whole, \hagn \, is a well-suited fiducial simulation because it has the approriate characteristics for the box size and the resolution to reproduce the Ly$\alpha$ forest and it is in agreement with observational galaxy properties, which is necessary to explore realistic feedback models.

\subsection{The fiducial Horizon-AGN simulation}
\label{sec:HAGN}
The cosmological hydrodynamical simulation \hagn \, is fully described in \cite{Dubois2014}, we present in this section the main relevant features. The simulation is run in a box of $\mathrm{L = \, 100\, Mpc.h^{-1}}$. It adopts a classical $\Lambda$CDM cosmology with total matter density $\mathrm{\Omega_m}$ = 0.272, dark energy density $\mathrm{\Omega_{\Lambda}}$ = 0.728,  amplitude of the matter power spectrum $\sigma_8$ = 0.81, baryon density $\Omega_b$ = 0.0455, Hubble constant $\mathrm{H_0}$ = 70.4 km $\mathrm{s^{-1}Mpc^{-1}}$ , and scalar spectra index $\mathrm{n_s}$ = 0.967, compatible with the WMAP 7 cosmology \citep{Komatsu2011}.
It contains $1024^{3}$ dark matter (DM) particles, which results in a DM mass resolution $\mathrm{M_{DM,res}} \sim 8 \times$ $10^7$ $\mathrm { M } _ { \odot }$, and initial gas-mass resolution of $\mathrm{M_{gas,res}} \sim 1 \times$ $10^7$ $\mathrm { M } _ { \odot }$.
It uses the adaptative mesh refinement code RAMSES \citep{Teyssier2002}. From the level 10 coarse grid, a cell is refined up to an effective resolution $\mathrm{\Delta x \sim 1 kpc} $ (level 17 at $z = 0$). Refinement is triggered in a quasi-Lagrangian manner: if the number of DM particles in a cell becomes greater than 8, or if the baryonic mass reaches eight times the initial baryonic mass resolution in a cell, a new level of refinement is triggered.

Gas cooling occurs by means of H and He cooling down to $10^4$ K with a contribution from metals following the model from \cite{Sutherland1993}. Reionization takes place after redshift $\mathrm{z_{reio}}$ = 10 due to heating from a uniform UV background from \cite{Haardt1996}.
The star formation is modeled with a Schmidt law $\dot{ \rho } _ { * } = \epsilon _ { * } \rho / t _ { \mathrm { ff } }$ with  $\dot { \rho } _ { * } $ the Star Formation Rate (SFR) density, $\mathrm{ \epsilon _ { * } = 0.02}$ the constant star formation efficiency and $ t _ { \mathrm { ff } }$ the local free-fall time of the gas \citep{Kennicutt1998, Krumholz2007}.
Star formation occurs only in cells with hydrogen gas density $\mathrm{\rho}$ exceeding $\mathrm{\rho_0 = 0.1 H.cm^{-3}}$ with a standard 2\% efficiency per free-fall time and follow the Schimdt-Kennicutt law \citep{Kennicutt1998}. Feedback from stellar winds, type Ia and type II supernovae are included to release mass, energy and metals in their surrounding environment assuming a Salpeter Initial Mass Function.

Black holes (BH) are represented by sink particles with an initial mass of $10^5$\msol. They can accrete gas in their surrounding environment at the Bondi-Hoyle-Lyttleton rate,
\begin{equation}
\dot { M } _ { \mathrm { BH } } = \frac { 4 \pi \alpha G ^ { 2 } M _ { \mathrm { BH } } ^ { 2 } \overline { \rho } } { \left( \overline { c } _ { s } ^ { 2 } + \overline { u } ^ { 2 } \right) ^ { 3 / 2 } },
\end{equation}
with $\alpha$ the dimensionless boost factor, $M _ { \mathrm { BH } } $ the BH mass, $\mathrm{\overline { \rho } }$ the mean gas density,  $\ \mathrm{\overline { c } _ { s }}$ the average sound speed and $ \mathrm{\overline { u } }$ the average gas velocity relative to the BH. $\mathrm{\alpha \geq 1}$ accounts for the lack of resolution in the accretion disk in star forming gas. We have $\alpha = \mathrm{(\rho / \rho_0)^2}$
if $\mathrm{\rho > \rho_0 }$ and $\alpha = 1$ otherwise. $\dot { M } _ { \mathrm { BH } }$ is limited by the Eddington accretion rate,
\begin{equation}
\dot { M } _ { \mathrm { Edd } } = \frac { 4 \pi G M _ { \mathrm { BH } } m _ { \mathrm { p } } } { \epsilon _ { \mathrm { r } } \sigma _ { \mathrm { T } } c },
\end{equation}
with $ \sigma _ { \mathrm { T }  }$ the Thomson cross-section, $c$ the speed of light and  $\epsilon _ { \mathrm { r } } = 0.1$ the radiative efficiency. AGN feedback injects a fraction $\epsilon_{\mathrm {f}}$ of the radiated energy in the medium in the form of kinetic and thermal energies. It implies,
\begin{align}
\mathrm{\Delta E}_{\mathrm{medium}} &= \epsilon_{\mathrm {f}} \mathrm {L} _ {\mathrm {r}} \\
&=  \epsilon_{\mathrm {f}}   \epsilon_{\mathrm {r}} \dot { M } _ { \mathrm { BH } }^2  c^2,
\label{eq:delta_e}
\end{align}
where $\mathrm {L} _ {\mathrm {r}} $ is the radiated energy. The feedbacks come in two modes \citep{Dubois2012}, depending on the value of the ratio of the accretion rate to its Eddington limit
\begin{equation}
\chi = \frac { \dot { M } _ { \mathrm { BH } } } { \dot { M } _ { \mathrm { Edd } } }.
\end{equation}

If $\chi > 10^{-2}$ the quasar mode is triggered as it is believed to happen mostly at high redshift, when the BH undergoes fast episode of growth. It is presumed that the BH emits large amounts of radiations that heat and ionize its environment. Therefore thermal energy is injected in a sphere of radius \ragn \,, by increasing the internal energy of the impacted gas cells with \ef \, $= 0.15$. \ragn \, is the radius of energy deposition.

If $\chi < 10^{-2}$ the radio mode is triggered. To account for the observed inflated cavities with strong magnetic fields, mass momentum and kinetic energy are injected in bipolar jets with \ef \, $= 1$. The jets are modeled as a cylindre of radius  \ragn \,  and height $2$ \ragn. \ragn \, is chosen to be $\Delta x$, the size of the smallest cell, after calibration to observations at $z = 0$.

\begin{figure}
  \centering
  \begin{minipage}{\columnwidth}
        \centering
        {\includegraphics[width=\columnwidth]{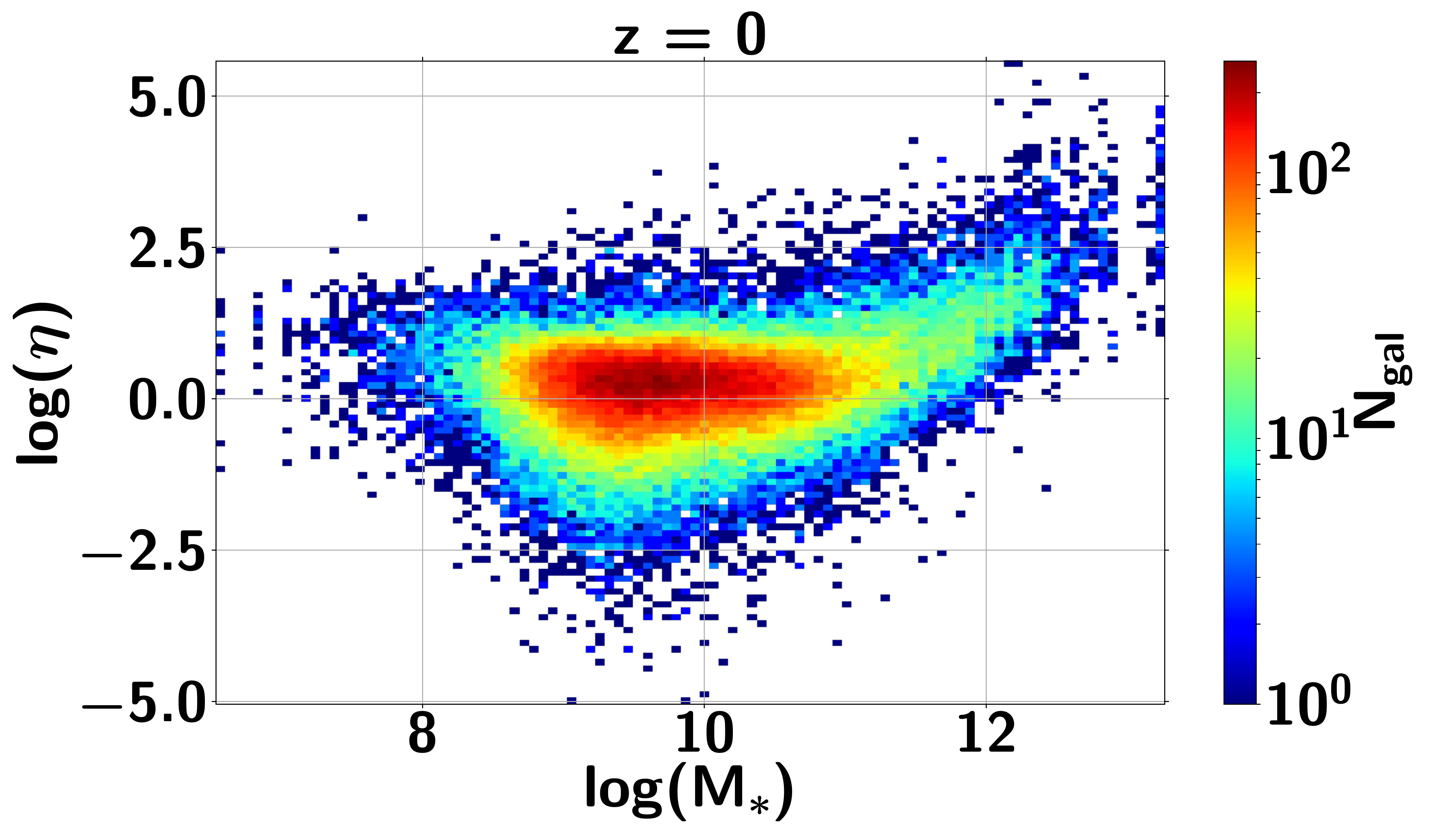}}
         {\includegraphics[width=\columnwidth]{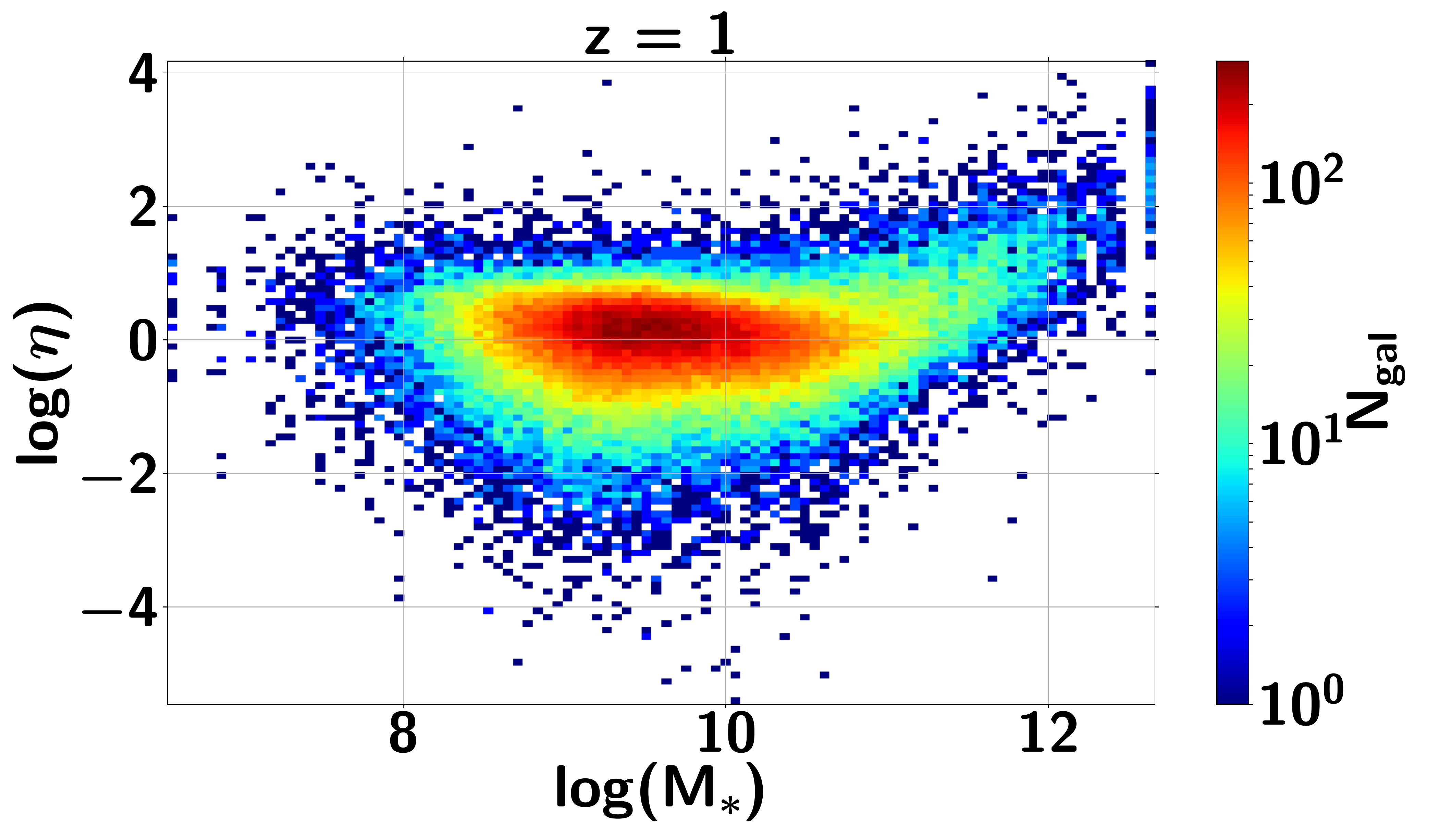}}
        {\includegraphics[width=\columnwidth]{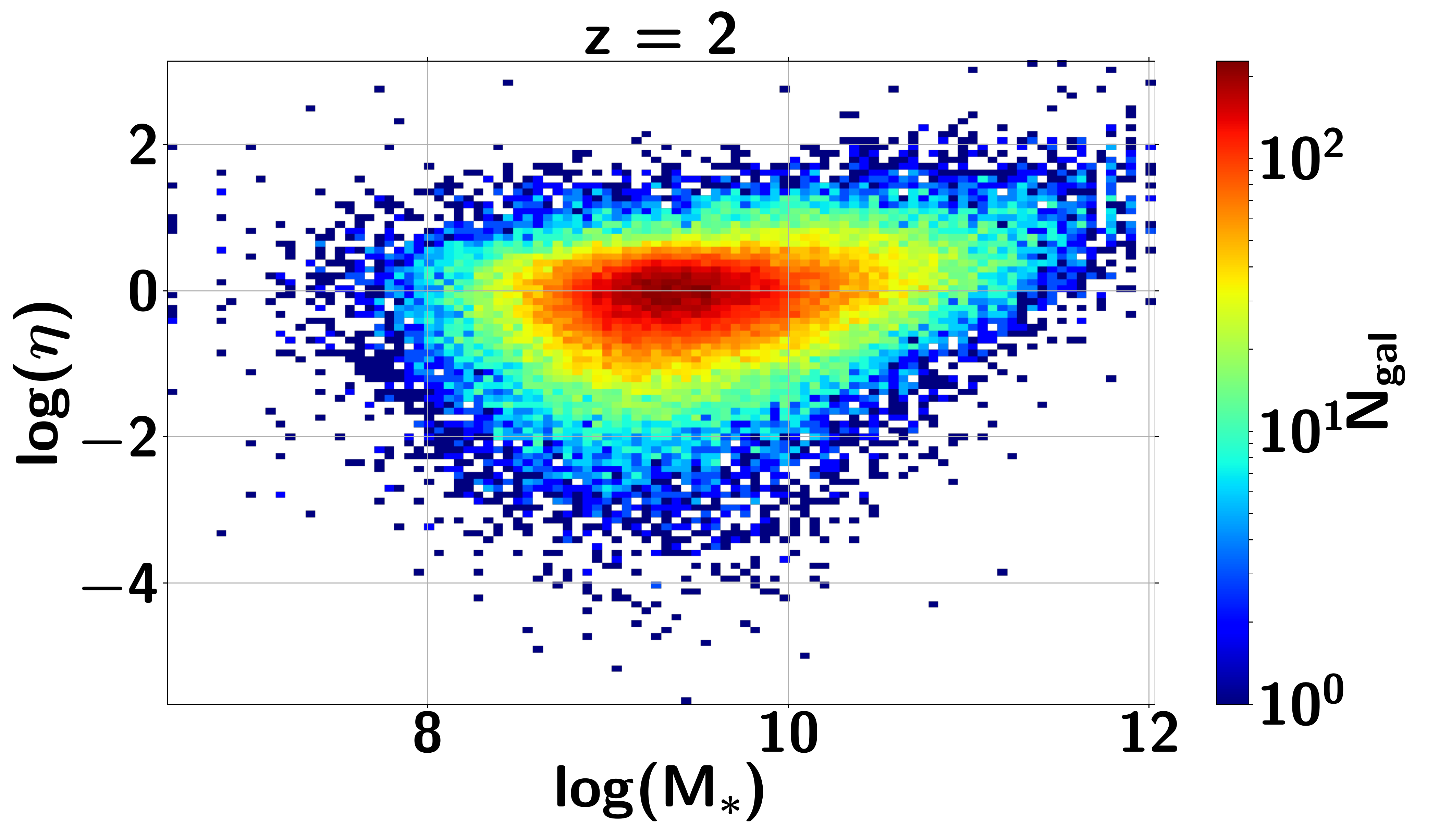}}
    \end{minipage}
    \caption{Mass loading factors of \hagn, $\eta$, in function of stellar mass $M_{*}$, at z = 0 (top), z = 1 (middle) and z = 2 (bottom). They are defined as $\eta = \dot { M } _ { \mathrm { outflows } } / \mathrm{SFR}$, where $\dot { M } _ { \mathrm { outflows } }$ is the mass outflow rate and $\mathrm{SFR}$ is the star formation rate.}
    \label{fig:load_factors}
\end{figure}

We expect AGN and stellar feedbacks to have different impacts; SN-driven winds are efficient at expelling gas mostly in low-mass halos because they are not fast enough to overcome the escape velocity of gravitational potentiel of high-mass halos \citep{Dekel1986}. AGN feedback are more efficient in high-mass halos because SN winds reduce BH growth in the central regions of galaxies by removing cold dense gas until the potential well is deep enough to confine the gas close to the BH \citep{Dubois2015,Habouzit2017}. Briefly, we expect a more homogeneous effect for SN feedback compared to AGN feedback. However, if both feedback mechanism happen at the same time we observe non-linear coupling, in the sense that dense cold gas of SN-driven winds is accelerated by hot outflows, powered by AGN, at much larger scales than without AGN \citep{Biernacki2018}. The aim of this study is to estimate the impact of AGN feedback on the 1D power spectrum of the \lya\ forest; the feedbacks coupling study is beyond the scope of the paper but it needs to be accurately calibrated. The mass loading factor is a key observable for the study of the \lya\ forest to validate this coupling calibration because it quantifies the amount of gas expeled by galactic feedback, which ultimatly strongly modifies the gas distribution of the IGM. We show in Fig.~\ref{fig:load_factors} the mass loading factors  $\eta$ in \hagn\, in function of stellar mass at z = 0, z = 1 and z = 2. These have been estimated using outflow rates and star formation rate measurements from \cite{Beckmann2017}. These tend to be $\eta \sim 1$ for galaxies with $M_{*} = 10^{10} \ \rm { M } _ { \odot }$ at $z = 1$ and $z = 2$ and tentatively increase toward lower redshifts. This is fully consistent with observations, e.g. see Fig.7 of \cite{Schroetter2019} or \cite{Schreiber2019}.

A companion simulation Horizon-noAGN (\hnagn) was run without AGN feedback. The same BH seeds are used so we do not have to account for shot noise. In addition, we ran restarts from $z=7$, \hlmin \, and \hnlmin, where we forced the refinement from level 10 to level 11 of the coarsest gas cells in order to perform convergence tests on the corrections. In other words, we increase the resolution in the less resolved regions where the diffuse Lyman-$\alpha$ forest gas belongs.

\subsection{The set of additional simulations: varying AGN feedback and feeding parameters}
\label{sec:variants}

To estimate uncertainties that encompasse the whole range of realistic AGN feedback models rather than relying on one single implementation, we performed six restarts from \hagn \, at redshift 7, when AGN feedback do not have noticeable effects yet. In the six additional simulations, we modify the three main subgrid feedback and feeding parameters that could impact the Ly$\alpha$ forest:
\begin{itemize}
  \item \clp \, and \clpp \, introduce stochasticity in the accretion rate in order to mimic the accretion of massive dense clouds in the interstellar medium, which are not captured by the HAGN resolution. In \cite{DeGraf2017}, the authors show that it can impact the evolution of the BH mass at high redshifts. \clp \, has a boost factor and Eddington limit ten times stronger 10\% of the time and a hundred time stronger 1\% of the time for \clpp. They are run to redshift 2.3 as we will show in Sec.~\ref{sec:results} that $\mathrm{\alpha}$ does not impact the corrections above the percent level.
  \item \rp \, and \hrm \, increase and decrease \ragn \, respectively, the other parameters are identical to those of \hagn. They are run to redshift 2.
  \item \ep \, and \hem \, increase and decrease \ef \, respectively. They are run to redshift 2.
\end{itemize}

The subgrid parameters for each simulation of the suite are summarized in Tab.~\ref{tab:vsimu}.
They were chosen so that the $\mathrm{M_{BH}}-\mathrm{M_*}$ relation and the mean fraction of gas $\mathrm{ f_{gas}}$ in galaxies span the observational uncertainties.

\begin{table*}
\centering

\begin{tabular}{|c | ccc|}
\hline
Simulation & $\alpha$ & \ragn \, & \ef \, \\
\hline
\hagn
&
$\left\{
\begin{array}{cc}
         (\rho / \rho_0)^2 & \textrm{if } \rho > \rho_0 \\
        1 & \textrm{otherwhise }
    \end{array}
    \right.$
 &
   $\Delta x$
   &
$\left\{
\begin{array}{cc}
        0.1 & \textrm{if } \chi < 10^{-2}  \\
        0.15 & \textrm{if } \chi > 10^{-2}
    \end{array}
    \right.$
 \\
\hline
\clp
&
 $\mathrm{10\%}$ of the time:   $\mathrm{10\alpha_\mathrm{{HAGN}}}$
 &
 $\mathrm {r} _ {\mathrm{AGN, HAGN}}$
 &
$\mathrm{\epsilon_{\mathrm{f,HAGN}}}$ \\
\clpp
&
 $\mathrm{1\%}$ of the time:   $\mathrm{100\alpha_\mathrm{{HAGN}}}$
&
$\mathrm {r} _ {\mathrm{AGN, HAGN}}$
&
$\mathrm{\epsilon_{\mathrm{f,HAGN}}}$ \\

\hline

\rp
&
 $\mathrm{\alpha_\mathrm{{HAGN}}}$
 &
  $2  \Delta x$
  &
$\mathrm{\epsilon_{\mathrm{f,HAGN}}}$
\\
\hrm
&
 $\mathrm{\alpha_\mathrm{{HAGN}}}$
&
 $0.5  \Delta x$
 &
$\mathrm{\epsilon_{\mathrm{f,HAGN}}}$\\

\hline

\ep
&
 $\mathrm{\alpha_\mathrm{{HAGN}}}$
 &
 $\mathrm {r} _ {\mathrm{AGN, HAGN}}$
 &
$\left\{
\begin{array}{cc}
        3 & \textrm{if } \chi < 10^{-2}  \\
        0.45 & \textrm{if } \chi > 10^{-2}
    \end{array}
    \right.$
    \\
\hem
&
  $\mathrm{\alpha_\mathrm{{HAGN}}}$
 &
  $\mathrm {r} _ {\mathrm{AGN, HAGN}}$
  &
$\left\{
\begin{array}{cc}
        0.33 & \textrm{if } \chi < 10^{-2}  \\
        0.05 & \textrm{if } \chi > 10^{-2}
    \end{array}
    \right.$ \\
\hline
\end{tabular}
\caption{Summary of the simulations used to estimate corrections and uncertainties due to the AGN feedback model. From left to right, the columns list: simulation name, value of the boost factor, the radius of energy deposition where $\Delta x$ is the smallest cell, and finally the energy efficiency. We stress that \ef \, can be superior to 1 as it represents the fraction of radiated energy injected in the medium, and not the fraction of total energy. }
\label{tab:vsimu}
\end{table*}

We take \hagn \, as the reference at all redshifts for $\mathrm{M_{BH}}-\mathrm{M_*}$ and $\mathrm{ f_{gas}}$  for the following reasons.  \cite{Dubois2014} calibrated the scaling relation at $\mathrm{z = 0}$ with observations. We also take \hagn \, as the reference in the scaling relation for highest redshifts given that observational uncertainties are very large, but also because its evolution is weak with increasing redshift following observations analysis \citep{Decarli2010, Merloni2010}, simulations \citep{Dubois2012,Sijacki2015} and analytical models \citep{Croton2006}. The mean fraction of gas appears to be in the realistic range measured by observations in our range of redshifts, even if no calibration were performed, as shown in Fig.~B2 of \cite{Welker2017}.

For both relations, we are interested in the deviations to the mean values, hence in the systematic uncertainties more than the intrinsic dispersion.
 For the mean fraction of gas, \cite{Tacconi2018}, give a total observational uncertainty of 0.2. However the intrinsic dispersion largely dominates over the systematic uncertainty and we choose to only take the contribution from this last term, i.e. we take  $\mathrm{\sigma _ { f} = 0.035}$. For the scaling relation, we choose to take the global uncertainties from \cite{Baron2019} for the slope and amplitude because the systematic term largely overcomes the dispersion. We then have $\mathrm{\sigma _ a} = 0.18$ and $\mathrm{\sigma _ b} = 0.13$, where  $\mathrm{\sigma _ a}$
 is the uncertainty on the slope and $\mathrm{\sigma _ b}$ the uncertainty on the amplitude.

\begin{figure*}
\includegraphics[width=\textwidth]{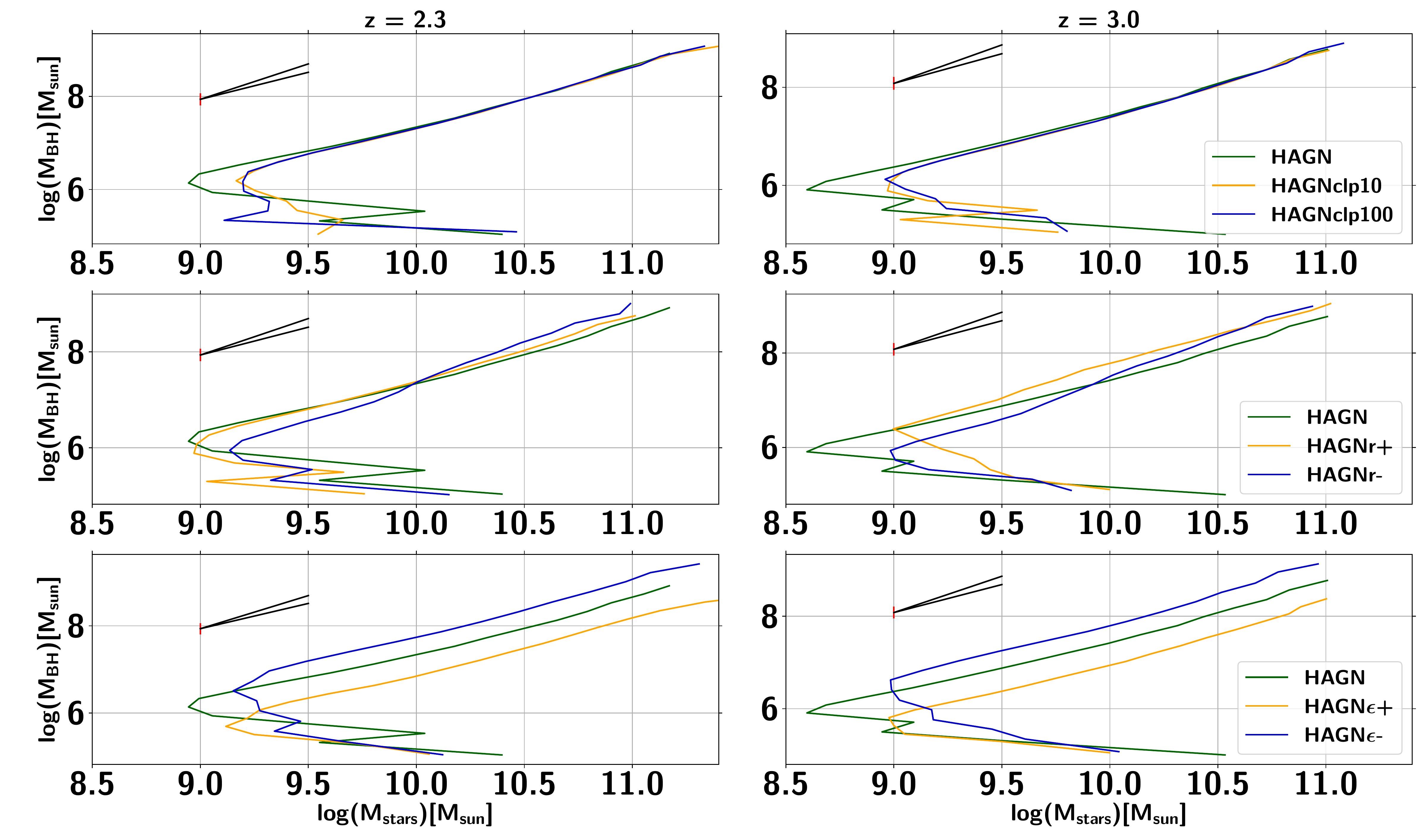}
\caption{$\mathrm{M_{BH}}-\mathrm{M_*}$ relations for the additional simulations compared to the fiducial simulation \hagn. It shows \clp \, and \clpp \, on the first line panels, \rp \, and \hrm \, on the second line panels and \ep \, and \hem \, on the third line panels. Left and right panels are at redshift 2.3  and 3.0 respectively. Observation uncertainties are overploted for $\mathrm{M_* = 10^{10.5} M_{\odot}}$, in red we show $\mathrm{\sigma_b}$ the uncertainty on the amplitude, and in black the two extremal slopes authorized by $\mathrm{\sigma_a}$}
\label{fig:magrel}
\end{figure*}

Fig.~\ref{fig:magrel} shows the scaling relations for the six additional simulations along with \hagn \, at redshifts 2.3 and 3.0. We represent the average value of the distribution of the stellar mass for a given bin of BH mass. Observation uncertainties from \cite{Baron2019} are overploted for $\mathrm{M_* = 10^{10.5} M_{\odot}}$. In red we show $\mathrm{\sigma_b}$ the uncertainty on the amplitude, and in black the two extremal slopes authorized by $\mathrm{\sigma_a}$.
Tab.~\ref{tab:linfits} gives the amplitude and slope of the linear fits with deviations to the reference model \hagn \, in terms of the observational uncertainties. We emphasize that we did the linear fits on the part of the relation where $\log \mathrm{(M_{BH}}) > 7 $, as there is not enough statistical power below this value, and on the top branch to be coherent with observations.

\begin{table*}
\centering
\begin{tabular}{c|cccc|cccc}
\hline
& $a _ {2.3}$ & $ b _ {2.3}$ & $\Delta a _ {2.3}$ & $\Delta b _ {2.3}$ & $a _ {3.0}$ & $b _ {3.0}$& $\Delta a _ {3.0}$ & $\Delta b _ {3.0}$\\
\hline
\hagn  & 1.34 & 8.63 & 0.0 & 0.0 & 1.39 & 8.77 & 0.0 & 0.0 \\
\hline
\clp  & 1.33 & 8.60 &$\mathrm{<\sigma _ a}$ & $\mathrm{<\sigma _ b}$& 1.40 &  8.76 &$\mathrm{<\sigma _ a}$ & $\mathrm{<\sigma _ b}$\\
\clpp & 1.34 & 8.62 &$\mathrm{<\sigma _ a}$ & $\mathrm{<\sigma _ b}$ & 1.43 & 8.78 &$\mathrm{<\sigma _ a}$ & $\mathrm{<\sigma _ b}$ \\
\hline
\rp  & 1.40 & 8.76 &$\mathrm{<\sigma _ a}$ & $\mathrm{\sigma _ b}$ & 1.20 &  8.99 &$\mathrm{\sigma _ a}$ & $\mathrm{1.7\sigma _ b}$\\
\hrm  & 1.59 & 8.98&$\mathrm{2\sigma _ a}$ & $\mathrm{2.7\sigma _ b}$ & 1.62 & 9.15 &$\mathrm{1.7\sigma _ a}$ & $\mathrm{2.9\sigma _ b}$ \\
\hline
\ep  & 1.36 & 8.17 &$\mathrm{<\sigma _ a}$ & $\mathrm{3.5\sigma _ b}$ & 1.40 & 8.32 &$\mathrm{<\sigma _ a}$ & $\mathrm{3.5\sigma _ b}$ \\
\hem  & 1.35  &  9.04 &$\mathrm{<\sigma _ a}$ & $\mathrm{3.15\sigma _ b}$ & 1.45 & 9.22 &$\mathrm{<\sigma _ a}$ & $\mathrm{3.5\sigma _ b}$\\
\hline
\end{tabular}
\caption{Slopes $a$ and amplitudes $b$ of the linear fits for the $\mathrm{M_{BH}}-\mathrm{M_*}$ relation of the additional simulations, such that $\mathrm{log\left(\frac{M_{BH}} {\mathrm{M } _ { \odot }}\right) = a.log\left( \frac{M_*} {10^{10.5}\mathrm{M } _ { \odot }}  \right) + b }$. The second and third columns are at redshift 2.3 and the sixth and seventh columns at redshift 3.0. It gives the deviation to the reference relation of \hagn \, with the $\mathrm{1\sigma}$ deviation being $\mathrm{\sigma _ a} = 0.18$ and $\mathrm{\sigma _ b} = 0.13$ at all redshifts.}
\label{tab:linfits}
\end{table*}

 Tab.~\ref{tab:fgas} gives the mean fraction of gas in the galaxies of the different simulations with deviations compared to \hagn. For the fraction of gas in a galaxy we take the mass of gas contained in a cylindre of radius of two times the effective radius and height 2 kpc oriented along the spin of the galaxy, with temperature $\mathrm{T < 10^6 K}$, to be coherent with the observations, and we compare it to the stellar mass contained in the same volume. We take the mean of all galaxies with mass $\mathrm{10^{10} M_{\odot} \leq M_{galaxy} \leq 10^{11} M_{\odot}}$ also to be coherent with observations.

 \begin{table}

 \centering
 \begin{tabular}{c|cc}
   \hline
   & $\mathrm{ f_{gas}}$ & $\mathrm{\Delta f_{gas}}$\\
\hline
\hagn  & 0.46 & 0.0 \\
\hline
\clp  & 0.45 & $\mathrm{<\sigma_f}$\\
\clpp & 0.42 &  $\mathrm{\sigma_f}$\\
\hline
\rp  & 0.36 & $\mathrm{3\sigma_f}$\\
\hrm  & 0.57 &  $\mathrm{2.7\sigma_f}$\\
\hline
\ep  & 0.38 & $\mathrm{2.3\sigma_f}$\\
\hem  & 0.56 &  $\mathrm{2.5\sigma_f}$\\
\hline
 \end{tabular}
 \caption{Mean fraction of gas $\mathrm{ f_{gas}}$ for the resimulations and the fiducial simulation. We also give the deviation to the reference relation of \hagn \, in term of the observational uncertainty $\mathrm{\sigma_f = 0.035}$ at $z=2$. }
  \label{tab:fgas}
 \end{table}

We can first notice that the evolution of the scaling relation between $\mathrm{z = 2.3}$ et $\mathrm{z = 3}$ is weak, as expected, for all the simulations, excepted for \rp \, and \hrm. On the first row of Fig.~\ref{fig:magrel}, \clp \, and \clpp \, appear to be almost identical to the top branch of \hagn. Tab.~\ref{tab:linfits} shows that the slopes and amplitudes are well below the $\mathrm{1\sigma}$ level at the two redshifts.
It confirms the study done in \cite{DeGraf2017} investigating the impact of stochasticity in the BH accretion rates. They show that clumpy accretion is significant at high redshifts for the BH evolution as it enables the accretion rate to outreach the Eddington limit. However they also show that the clumpy and not clumpy accretion models converge around redshift 6. Our simulations confirm that, at our resolution and our redshift range, the clumpy accretion does not have noticeable effect on the Maggorian relation. We only see a $\mathrm{1\sigma_f}$ effect for \clpp \, on the mean fraction of gas \footnote{If the stochasticity does not seem to have a strong impact on the mean fraction of gas in galaxies neither on the  $\mathrm{M_{BH}-M_*}$ relation, it however affects the total mass of the galaxy by decreasing both the mass of gas and stars. We observe galaxies in \clpp \, about two times less massive than in \hagn. In that sense, the relations $\mathrm{M_{BH}-M_{*}}$ for \clp \, and \clpp \, appear to be at the limit of the observational uncertainty.}.

\rp \, is not very well constrained by the scaling relation, we mostly observe an increase in the fitted amplitude at high redshift. It is in agreement with \cite{Dubois2012}, increasing \ragn \, leads to higher $\mathrm{M_{BH}}$ because the feedback is less energetic in the medium surrounding the BH. Therefore less gas is ejected and accreted yielding to a lower fraction of gas in galaxies. Indeed Tab.~\ref{tab:fgas} shows a mean fraction of gas at $\mathrm{3\sigma_f}$ for \rp.
By lowering \ragn \, below the cell size, we inject more energy to smaller gas mass which is ejected even further, but less gas is affected. The galaxies in \hrm \, contain more gas than in \hagn \, at about $\mathrm{2.7\sigma_f}$, because the feedback is stronger than in the first case, to self-regulate its growth the BH accrete less gas.

\ep \, and \hem \, slopes of the $\mathrm{M_{BH}}-\mathrm{M_*}$ relation are totally in agreement with \hagn \, at the two considered redshifts. However, the higher the efficiency the less compatible is the fitted amplitude. Our results are again consistent with the study done in \cite{Dubois2012}. At a given galaxy mass bin, more massive BHs are obtained if we decrease the efficiency. Following Eq.~\ref{eq:delta_e}, if the efficiency is decreased, the BH counterbalances by being more massive and accreting more gas in order to inject the same total amount of energy in the medium and self-regulate its growth. The deviation of the amplitude in the scaling relation to \hagn \, is constant with redshift, and is more than $\mathrm{3\sigma}$.
The two simulations are less constrained by $\mathrm{ f_{gas}}$ with deviations around $\mathrm{2.5\sigma_f}$. Nevertheless, this is consistent with \cite{Dubois2012} as less efficiency in the feedback leads to larger accretion rates hence to less gas in the galaxy.

Our set of additional simulations varying the main feedback parameters appear to largely deviate, i.e. at more than $\mathrm{3\sigma}$ in terms of observational uncertainty, from at least one of the chosen observables for \ragn \, and \ef, and at $\mathrm{1\sigma}$ for $\mathrm{\alpha}$. By spanning the observational uncertainties  we show that we cover the whole range of probable feedback models.

%% file: p1d.tex
We present in Sec.\ref{sec:method} the numerical methods used to estimate the 1D power spectra of the flux fluctuations in the \lya \, forest, and we present in Sec.\ref{sec:lya} its main characteristics in the Horizon-AGN simulation.

\subsection{Numerical methods}
\label{sec:method}

The flux fluctuations in the \lya \, forest along lines of sight (LOS) at redshift $z$, are well described by the 1D power spectrum (\plya). It is defined as the Fourier Transform of the flux density contrast $\delta _ \phi$, where
\begin{equation}
  \label{eq:delta}
  \delta _ {\phi} = \frac{\phi }{\left\langle \phi(z)  \right\rangle} - 1
\end{equation}
with $\phi$ the flux and $\left\langle \phi(z)  \right\rangle$ the mean flux at redshift $z$.

The computation of the transmitted flux fraction requires the knowledge of the mass, density, temperature at each point of the box. We choose to use SPH equations to perform this 3D mapping for the following reasons. First, it is too strong an assumption to consider that these scalar fields are constants in the AMR cells. Then, the state of an AMR cell influences its neighbors. We want to parametrize the fields in the box as smooth functions and not as unrealistic step-functions.  To do so, we transform the AMR gas cells into particles. We loop over each gas cell and place a particle with the total mass of the cell at its center using the rdramses tool \footnote{\url{http://www.astro.lu.se/~florent/rdramses.php}}. We use the 3D cubic spline kernel introduced in \cite{Monaghan1985} to smoothly distribute the quantities of interests of each particles over its neighboring cells:
\begin{equation}
  \label{eq:kernel}
  W \left( q _ { j } \right) = \left\{ \begin{array} { l l } { \left[ 1 + q _ { j } ^ { 2 } \left( - 1.5 + 0.75 q _ { j } \right) \right] \cdot \frac { 1 } { \pi } } & { \left| q _ { j } \right| \leq 1 } \\ { \left[ 0.25 \left( 2 - q _ { j } \right) ^ { 3 } \right] \cdot \frac { 1 } { \pi } } & { 1 < \left| q _ { j } \right| \leq 2 } \\ { 0 } & { \left| q _ { j } \right| \geq 2 } \end{array} \right.
\end{equation}
where $q _ { j } = \left| \mathbf { r } - \mathbf { r } _ { \mathrm { j } } \right| / h _ { j }$ is the reduced distance to particle $i$.
The smoothing length $h$ is chosen such that the volume inside the sphere of radius $h$ is equal to the volume within the considered cubic cell. Then,
\begin{equation}
  \label{eq:smooth}
  \mathrm{\left[\frac{L_{box}}{2^{l}}\right]^3 = \frac{4}{3} \times \Pi  \times h^3},
\end{equation}
with $l$ being the level of the cell. The simulation does not go below level 15 for redshift above 2. Finally, we can derive each scalar fields at every points of the box using the following SPH equation:
\begin{equation}
  \label{eq:sph}
  A ( \mathbf { r } ) = \sum _ { j } m _ { j } \frac { A _ { j } } {\mathrm{ N _ {H0, j } }} W \left( \left| \mathbf { r } - \mathbf { r } _ { j } \right| , h _ { j } \right),
\end{equation}
where A one of the scalar quantity, $ \mathbf { r }$  a position in the cube, $h$ the smoothing length and $W$ the kernel functions described in Eq.\ref{eq:kernel}, and finally $\mathrm{N_{H0}}= \frac{n_{H0}}{n_H}$ is the neutral fraction of hydrogen. The index $j$ loops over all the gas particles in the simulation box.

The neutral hydrogen fraction is fundamental for the computation of \plya. To model the chemistry of the gas, we consider the IGM as having the primordial gas abundances with hydrogen abundance $X = 0.76$, and helium abundance $Y = 0.24$. This is in agreement with the recent CMB observations \citep{Planck2018Cosmo}. $\mathrm{N_{H0}}$ is estimated following the classical hypothesis that the IGM gas is optically thin and in ionization equilibrium but not in thermal equilibrium. We only consider collisional ionization cooling, radiative recombination cooling and photo-heating from a uniform UV background to impact the chemical evolution of the 6 atomic species $\mathrm{H_0}$, $\mathrm{H^+}$, $\mathrm{He0}$, $\mathrm{He^+}$, $\mathrm{He^{2+}}$ and $\mathrm{e^-}$.
It leads to the following set of equations:
\begin{align}
  \mathrm{n_{H0} = n_H \alpha_{H^+}/(\alpha_{H^+} + \Gamma_{e,H0} + \frac{\Gamma_{\gamma,H0}}{n_e})} \\
  \mathrm{n_{H^+} = n_e - n_{H0} }\\
  \mathrm{n_{He^+} = (n_{He0} \alpha_{He^+})/(\Gamma_{e,He0} + \frac{\Gamma_{\gamma,He0}}{n_e})} \\
  \mathrm{n_{He^+} = Y n_H/ (1 + \frac{\alpha_{He^+}}{\Gamma_{e,He0}+\frac{\Gamma_{\gamma,He0}}{n_e}} + \frac{\Gamma_{e,He^+} + \frac{\Gamma_{\gamma,He^+}}{n_e}}{\alpha_{He^{2+}}}) }\\
  \mathrm{n_{He^{2+}} = n_{He^+} (\Gamma_{e,He^+} + \frac{\Gamma_{\gamma,He^+}}{n_e})/(\alpha_{He^{2+}}) }\\
  \mathrm{n_e = n_{H^+} + n_{He^+} + n_{He^{2+}}}, \\
  \label{eq:ioneq}
\end{align}
with $\alpha$ the recombination rate, $\mathrm{\Gamma_e}$ the collisional cooling rates and $\mathrm{\Gamma_\gamma}$ the photoionization rates.
If we consider that helium is fully ionized either once or twice, and if we neglect the other ionization state, then the electron fraction is only function of the hydrogen density and the neutral hydrogen fraction from Eq.\ref{eq:ioneq} can be easily computed. We have $\mathrm{n_e} 	\sim 1.15 \mathrm{n_H}$ and $\mathrm{n_e} 	\sim \mathrm{1.10 n_H}$ for the full first and second ionization respectively. We checked that making the assumption that Helium is either once or twice ionized  does not significantly change the corrections with differences at the level of $\mathrm{10^{-3}}$. In the following we make the calculations of the free electron fraction considering that Helium is fully ionized twice. The set of equation reduces to
\begin{equation}
  \mathrm{n_{H0} = n_H \alpha_{H^+}/(\alpha_{H^+} + \Gamma_{e,H0} + \frac{\Gamma_{\gamma,H0}}{1.15n_H})}.
\end{equation}
 We use the radiative cooling rates from \cite{Abel1997}, the collisional cooling rates from \cite{Katz1995} and the photoionization rates from \cite{Theuns1998}.

Once all the required fields are computed for each gas particle, we extract $\mathrm{50~000}$ LOS  parallel to one of the axis of the box (which is not the same for all LOS), and  whose origin and axis are randomly drawn, following the traditional procedure \citep{Croft2002, Gnedin2001}.

We divide the spectra in $\mathrm{N_{bin}=2048}$ bins with coordinate $x$(Mpc/h) in real space and $u$(km/s) in velocity space, such that $u(1+z) = x H(z)$. For each pixel $j$ of each LOS, we use the SPH equation of Eq.~\ref{eq:sph} to derive the density $\mathrm{n_{H,j}}$, the temperature $T_j$ and the peculiar velocity $v_j$ of the gas in this pixel. The observed velocity is then,
\begin{equation}
  v_{obs,j} = v_{j//} + \frac{H(z)}{1+z}x,
\end{equation}
  where $v_{j//}$ is the peculiar velocity of the gas along the LOS and $x$ the pixel coordinate in real space. From this we estimate the optical depth $\tau$ for H0  using an analytic approximation to the Voigt-Hjerting function, with which Voigt-profile are modeled following \cite{Tepper2006}. In velocity space, peculiar velocities modify the optical depth by shifting the absorption positions and broadening the lines \cite{McDonald2003}. We thus have
  \begin{equation}
     \tau_s(u) = \int_0^{L/2} \tau(x') \frac{1}{\sqrt{2\pi}b(x')}\exp\left(-\left(\frac{u-v_{obs}(x')}{b(x')}\right)^2\right)dx',
  \end{equation}
where $\tau_s(u)$ is the optical depth in redshift space at velocity coordinate $u$,  $\tau(x')$ the optical depth in real space at spatial coordinate $x'$ and $b(x')=\sqrt{2k_BT(x')/m_H}$ is the Doppler parameter with $k_B$ the Boltzmann constant and $m_H$ the mass of the hydrogen atom. All the \plya \, computations in the following are done in redshift space.

We highlight the fact that, on the contrary of most of the hydrodynamical simulations working with the \lya \, forest, we do not rescale the optical depths such that the mean flux $\left\langle \phi(z)  \right\rangle$ match the observations. We are interested in differences due to AGN feedback and it can include differences in the mean flux. Moreover there are no reasons that \hagn ~ and \hnagn ~ should have the same mean flux as they do not represent the same universe.
Then we can compute at each pixel $j$ the flux density contrast $\delta _ {\phi,j}$ where the flux is $\phi _j = e^{-\tau_j}$, and the mean flux $\left\langle \phi(z)  \right\rangle$ is estimated from the ensemble of the pixels along all LOS. Finally, the 1D power spectrum, \plya \, is constructed by taking the Fourier Transform of the transmitted flux fraction field using a Fast Fourier Transform (FFT) algorithm.

\subsection{The Ly$\alpha$ forest in Horizon-AGN}
\label{sec:lya}

\begin{figure}
\begin{center}
\includegraphics[width=\columnwidth]{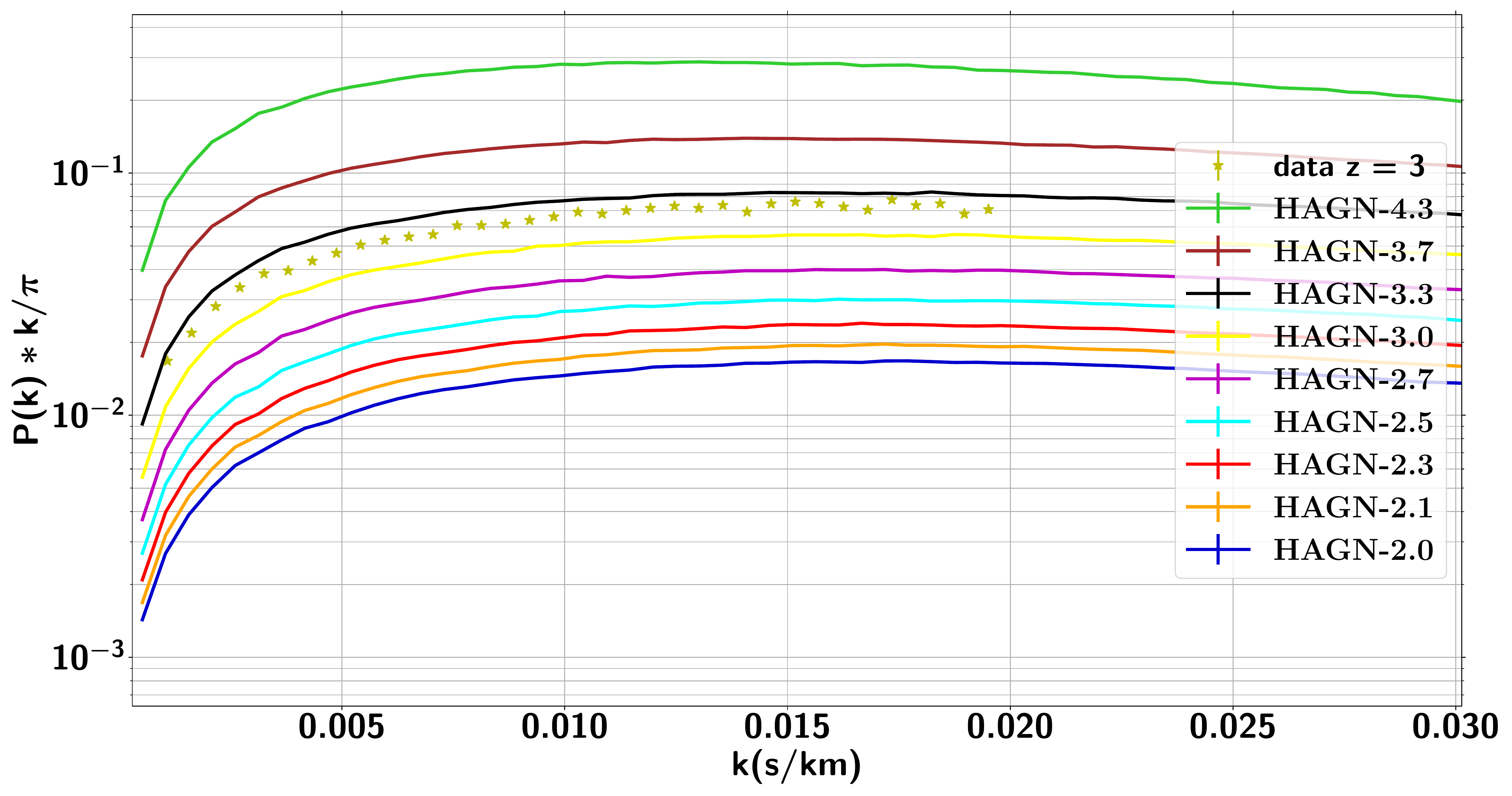}
\caption{Comparison of the \plya \, of \hagn \, at different redshifts in plain lines with the eBOSS DR14 data at redshift 3.0 for the yellow stars. Error bars represent the statistical error on the 50 000 LOS for the simulation \plya, and the combination of statistical and systematics uncertainties derived in \protect\cite{Chabanier2019a} for the observational data.}
  \label{fig:data}
\end{center}
\end{figure}

We show in Fig.~\ref{fig:data} the evolution with redshift of the \plya \, for \hagn \, in plain lines. The error bars represent the root-mean-square of the 50 000 LOS sample, they are well below the percent level at all redshifts. The yellow stars are the \plya \, data points that were derived using the eBOSS DR14 release \citep{Abolfathi2018} in  \cite{Chabanier2019a} at redshift 3, we include statistical and systematics uncertainties in the error bars. Simulations and observations are in  broad agreement, both in shape and amplitude. We do not require better agreement, as we are only interested in differences produced by AGN feedback.

\begin{figure}
\begin{center}
\includegraphics[width=\columnwidth]{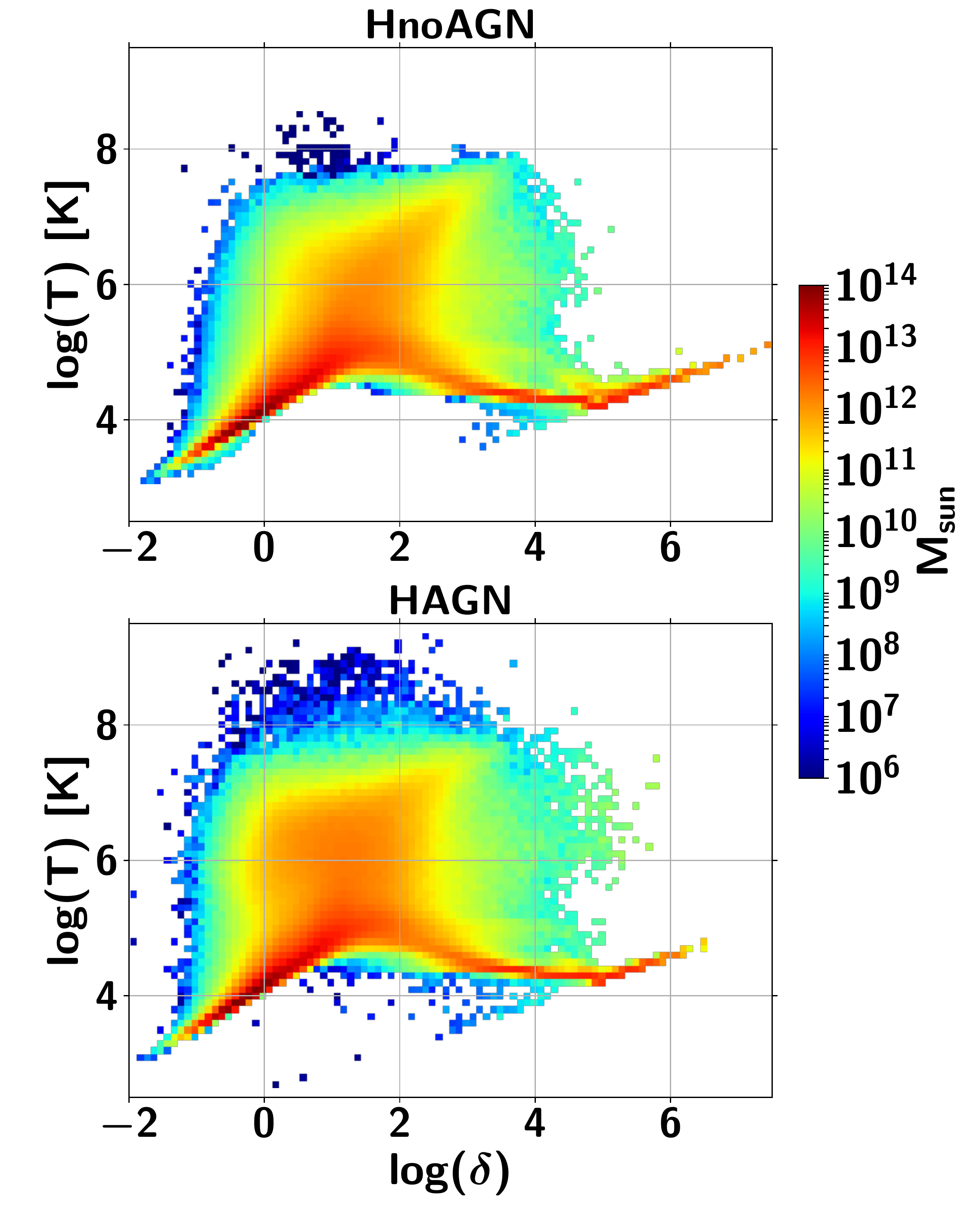}
\caption{Mass-weighted temperature-density diagram of the gas cells in \hagn ~ (bottom) and \hnagn ~ (top) at redshift 2.0 in logarithmic scales. The temperature $T$ is in Kelvin, and $\delta $ is the density contrast $\delta = \rho / \overline \rho -1$. }
  \label{fig:hist2d}
\end{center}
\end{figure}

Fig.~\ref{fig:hist2d} presents the mass-weighted temperature-density diagram of \hagn  ~ and \hnagn ~ at redshift $z = 2$ in logarithmic scales. The four populations constituing the baryonic gas are clearly visible; the cold diffuse density IGM, the hot IGM, the hot high density virialized gas from clusters and finally, the cold condensed star forming gas. The cold IGM phase constitutes the \lya \, forest we are interested in, it contains a very large fraction of the baryonic gas, both in volume and mass, and follows a linear relation between $\mathrm{log(T)}$ and $\mathrm{log(\rho)}$, as seen in observations and other cosmological hydrodynamical simulations including cooling \citep{Lukic2014,Borde2014}. The Jeans polytrope is also clearly visible at high densities, for $\rho > \rho_0 = 0.1 \mathrm{H.cm^{-3}} $, with the following EoS,
\begin{equation}
\label{eq:eos}
\mathrm{T = T_0 \left(\frac{\rho}{\rho_0}\right)^{p-1}},
\end{equation}
where $p = 4/3$ is the polytropic index of the gas. The reason to artificially increase the temperature of condensed star forming gas in the simulation is twofold. First, to increase the Jeans' lenght and avoid numerical artificial instabilities \citep{Truelove1997}, but also to account for the thermal heating of the ISM by SNe explosions \citep{Springel2005b}. However, as the neutral fraction $\mathrm{N_{H0}}$ is greatly dependent on the temperature so is the \plya. Hence, we checked that taking $\mathrm{T = 10^4K}$  in post-processing instead of keeping the artificially enhanced temperature from the simulation, for all gas cells with $\mathrm{\rho > \rho_0}$ do not change the results. Of course, modifying the temperature in such dense regions does not impact the \plya \, that dominantly probes the very diffuse gas. We will come back to the comparaisons of the two diagrams in Sec.~\ref{sec:results}

%% file: results.tex
\subsection{Impact of AGN feedback on the \plya}
\label{sec:correction}
\begin{figure}
 \begin{center}
 \includegraphics[width=\columnwidth]{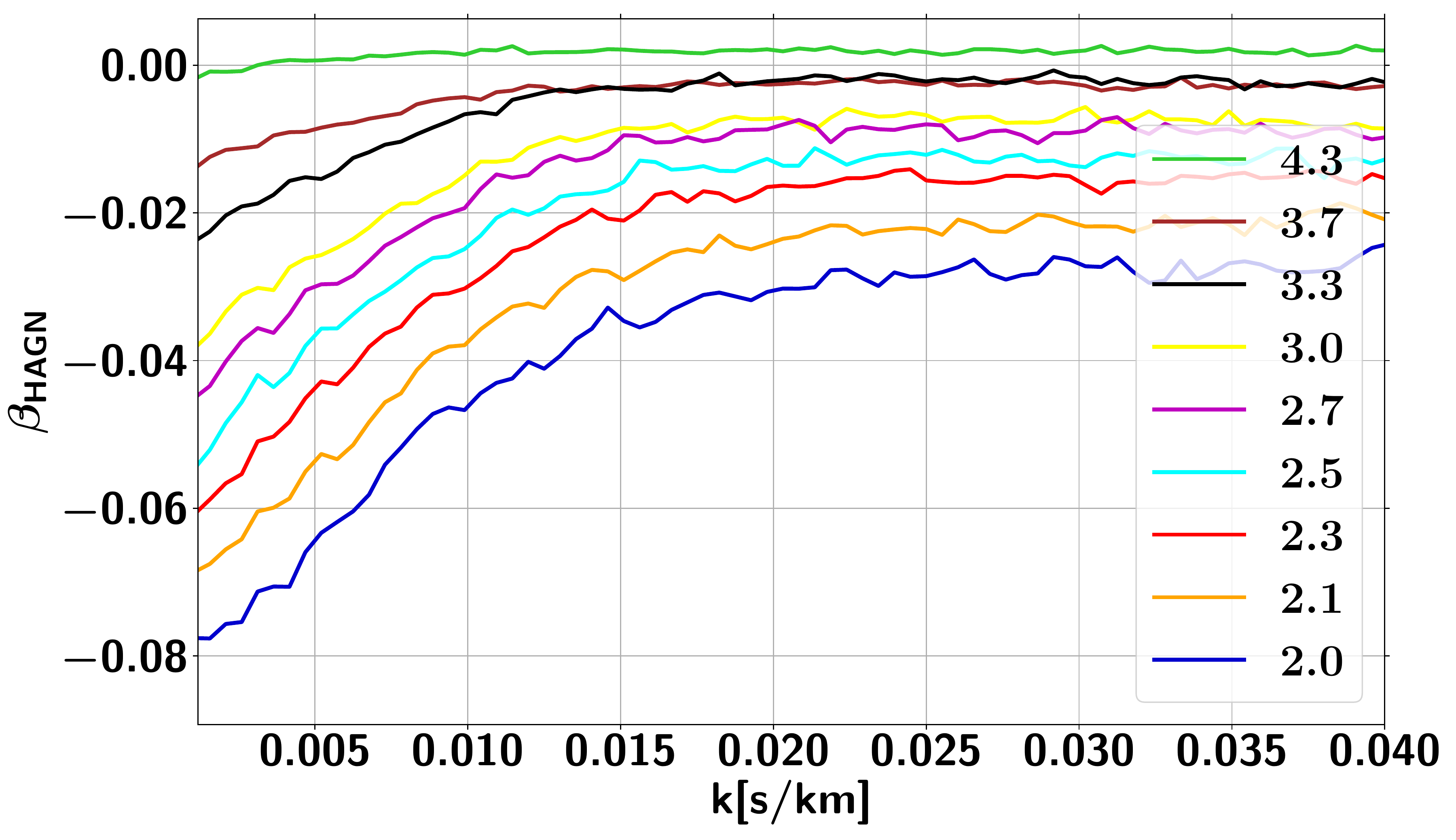}
 \caption{Correction $\beta_{\mathrm{HAGN}}$ of AGN feedback using the fiducial simulation \hagn. The different lines are for the nine different redshifts from $z=4.25$ to $z=2$.}
   \label{fig:hagn_corr}
 \end{center}
 \end{figure}

Fig.~\ref{fig:hagn_corr} shows the corrections $\beta$ estimated from the fiducial simulation \hagn. We take the correction due to AGN feedback as the deviation to one of the ratio of the \plya \, in \hagn \, to those in \hnagn \, using the same $\mathrm{50\,000}$ LOS, such that,
\begin{equation}
\frac{\mathrm{P_{Ly\alpha}(HAGN)}}{\mathrm{P_{Ly\alpha}(HnoAGN)}} = 1 + \beta.
\label{eq:correction}
\end{equation}
 The results are displayed at different redshifts from $z=4.25$ to $z=2.0$. We observe a suppression of power that increases with decreasing redshifts and increasing scales. The enhancement of suppression of power at large scales is already noticeable at $z=4.25$ and rises from less than 1\% to 8\% at $z=2.0$.

\begin{figure}
\begin{center}
\includegraphics[width=\columnwidth]{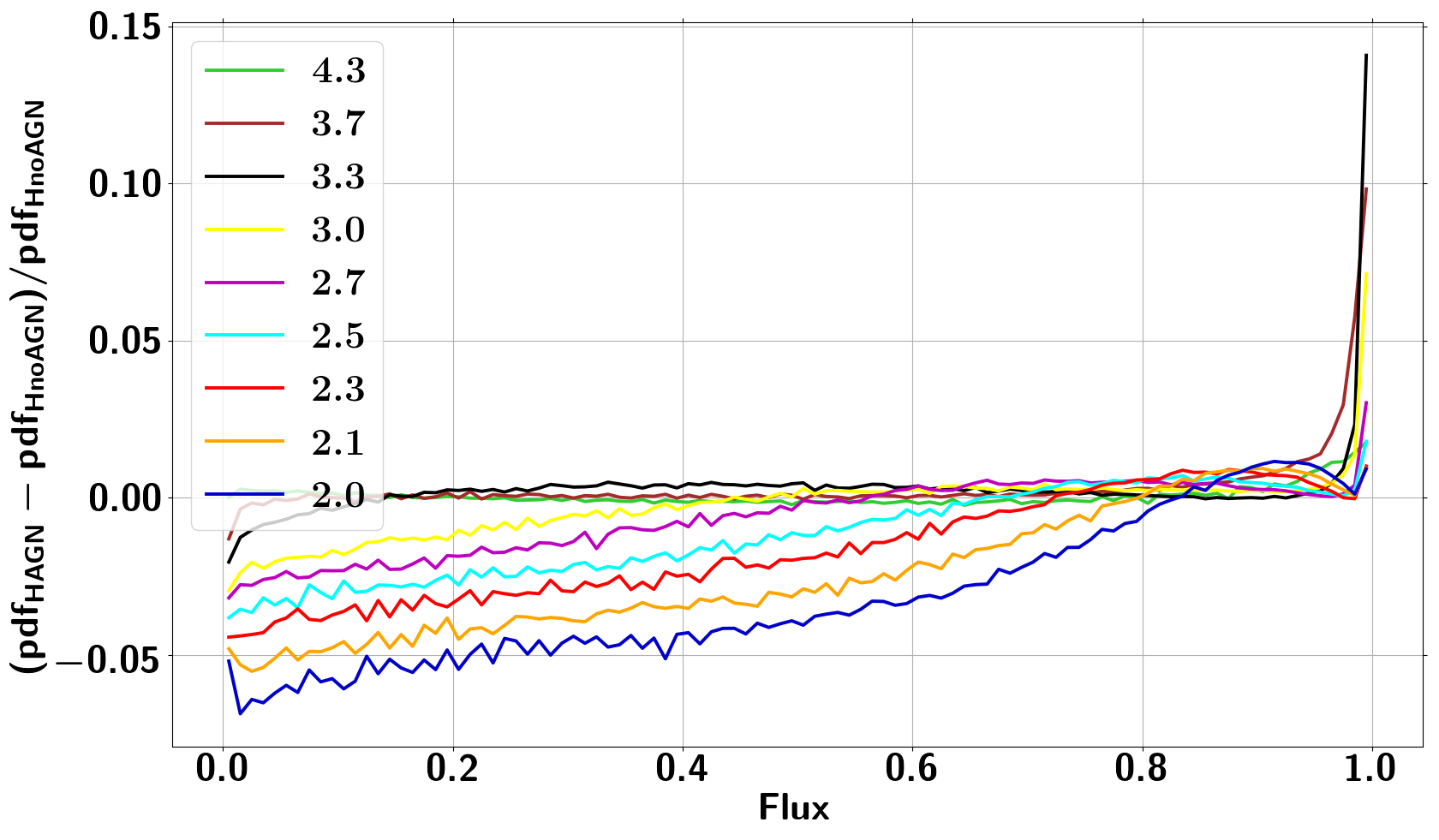}
\caption{Differences in the flux probability distribution betwen \hagn\ and \hnagn\ at all redshifts using pixels from the 50 000 LOS sample.}
  \label{fig:pdf}
\end{center}
\end{figure}

\begin{figure*}
    \centering
    \begin{minipage}{\textwidth}
        \includegraphics[width=0.985\textwidth]{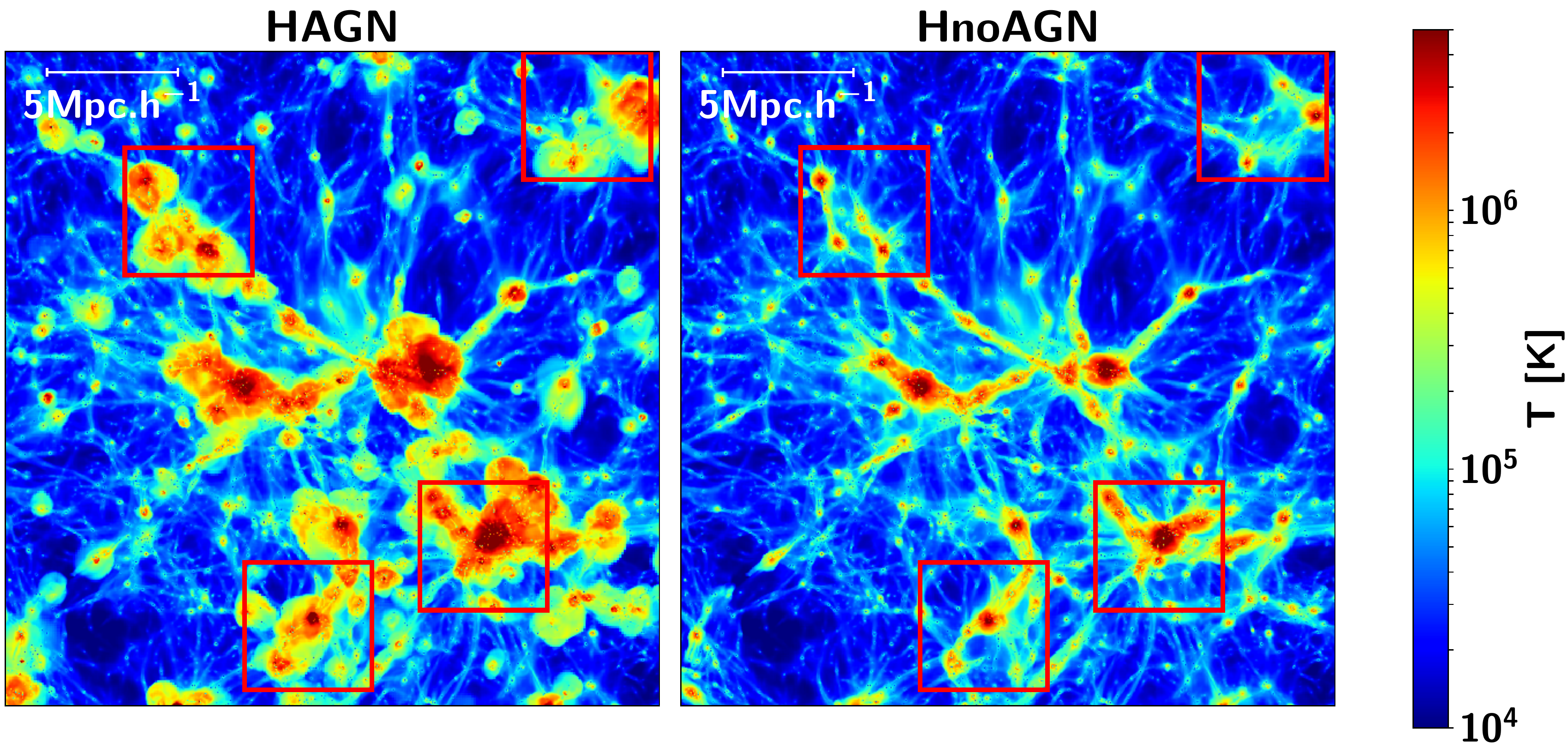}
        \includegraphics[width=\textwidth]{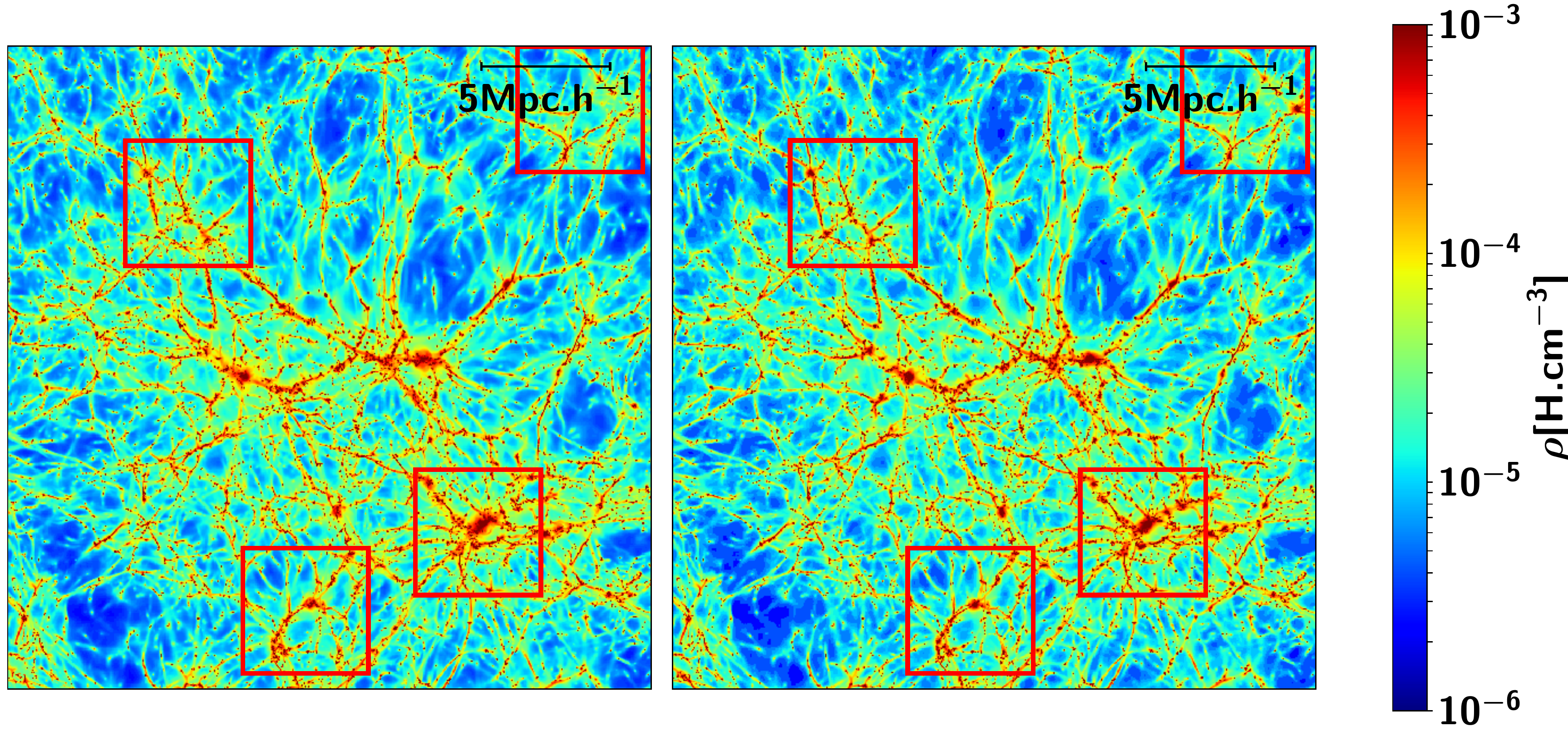}
     \end{minipage}
    \caption{Projected temperature (top) and density (bottom) maps of \hagn \, on the left and \hnagn \, on the right at $z = 2.0$ encoded in $\mathrm{\log(T)}$ and $\mathrm{\log(\rho)}$ unit. Boxes are 25 $\mathrm{Mpc.h^{-1}}$ in comoving coordinate. The four red boxes are regions A, B, C and D from left to right and top to bottom, we show the zoom of these regions on Fig.~\ref{fig:Tr_all}}
    \label{fig:Trmap}
\end{figure*}

As previously said in Sec.~\ref{sec:method} we do not rescale the mean flux for \hagn \, and \hnagn \, and we observe a global decrease of power, i.e. an increase of the mean flux, with AGN feedback. As shown in Fig~\ref{fig:pdf}, we observe a strong decrease in the number of pixels with low-flux transmittivity. It reflects the combination of a net increase of temperature, ionizing the ambient medium, but also the redistribution of gas from small to large scales.
The gas heating is clearly visible on the projected temperature maps of \hagn \, and \hnagn \, in the top panels of Fig.~\ref{fig:Trmap}. The left panel (\hagn \,) displays hotter bubbles than the right panel (\hnagn \,), that extend to larger scales and reach the IGM. This is in agreement with the temperature-density diagrams of Fig.~\ref{fig:hist2d} where we observe more pixels in the diffuse region, i.e. $\log(\delta) < 2$ with temperature  $\mathrm{T >10^5 K}$ in addition of the appearance of pixels with  $\mathrm{T >10^8 K}$ with AGN feedback. Indeed the hot IGM contains 18\% of the mass in \hagn \, and 12\% in \hnagn. The temperature is also higher in the dense region, i.e. $\log(\delta) > 2$, but the heating is less efficient as the temperatures do not go above $\mathrm{10^7}$/$\mathrm{10^8 K}$. This net increase of temperature is due to the injected thermal energy of quasar mode black holes that dominates compared to radio mode in our redshift range and ionizes the surrounding gas.
Because the \plya \, probes neutral hydrogen and because there is more ionized gas the power spectrum exhibits a suppression of power at all scales.

\begin{figure*}
  \centering
  \begin{minipage}{\textwidth}
        \centering
        {\includegraphics[width=0.5\textwidth,  height=3.5cm,keepaspectratio]{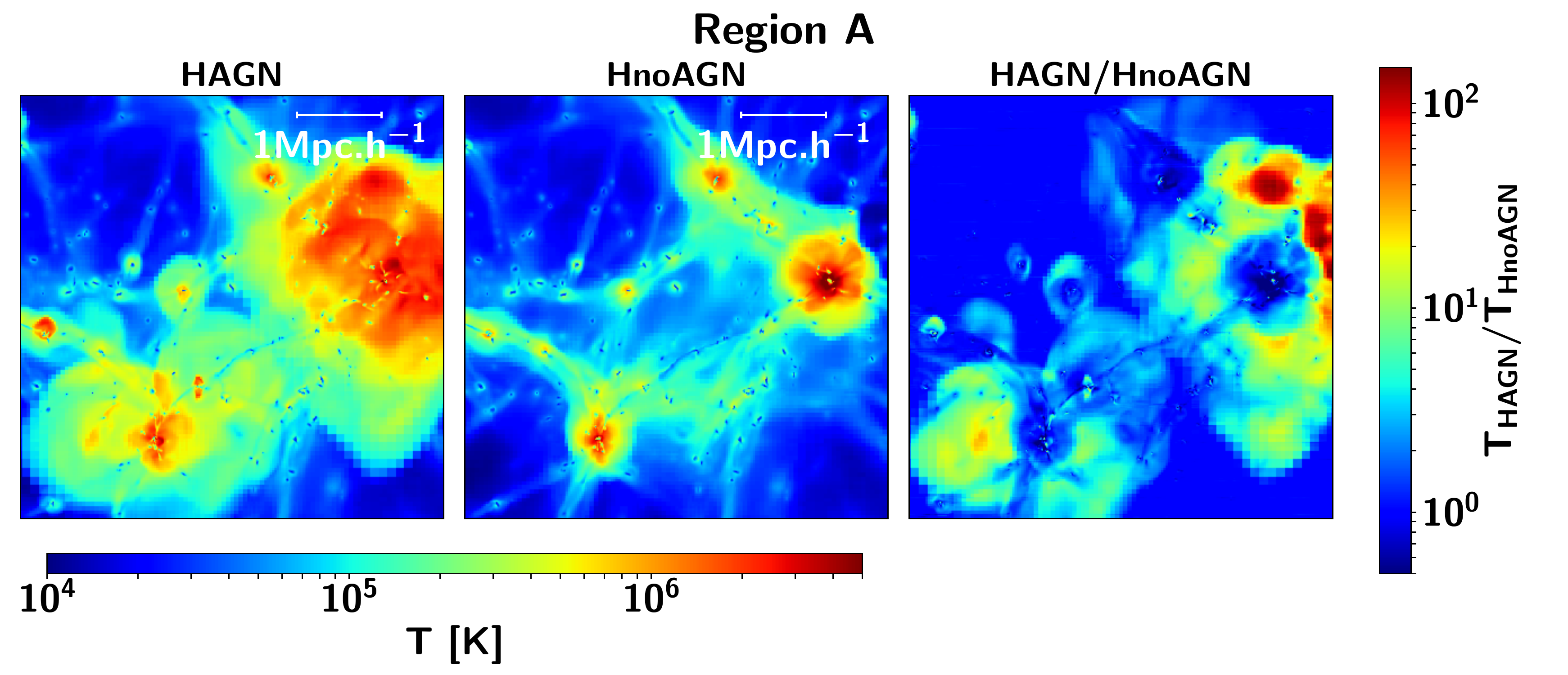}}
        \qquad
        \vspace{-1.5mm}
         {\includegraphics[width=0.5\textwidth,  height=3.5cm,keepaspectratio]{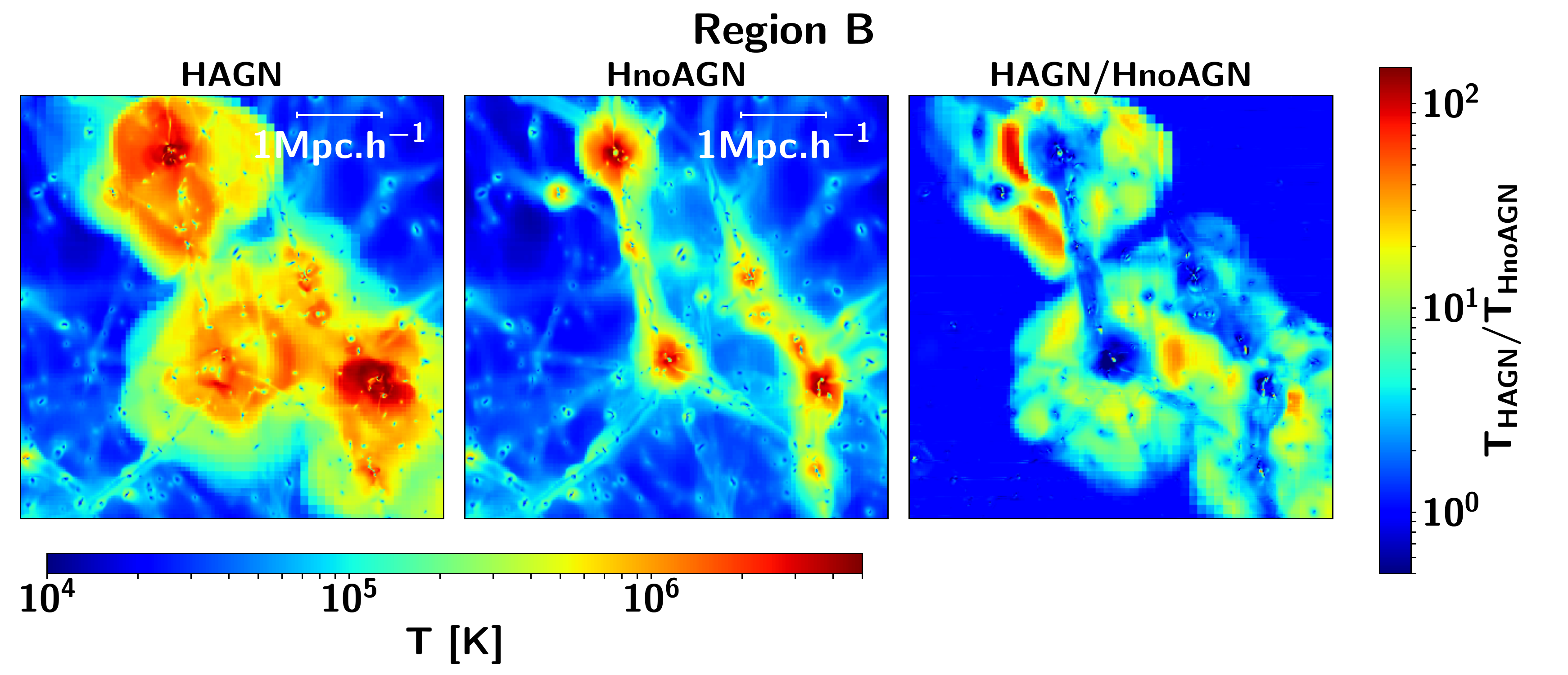}}
          \vspace{8mm}
        {\includegraphics[width=0.5\textwidth,  height=3.5cm,keepaspectratio]{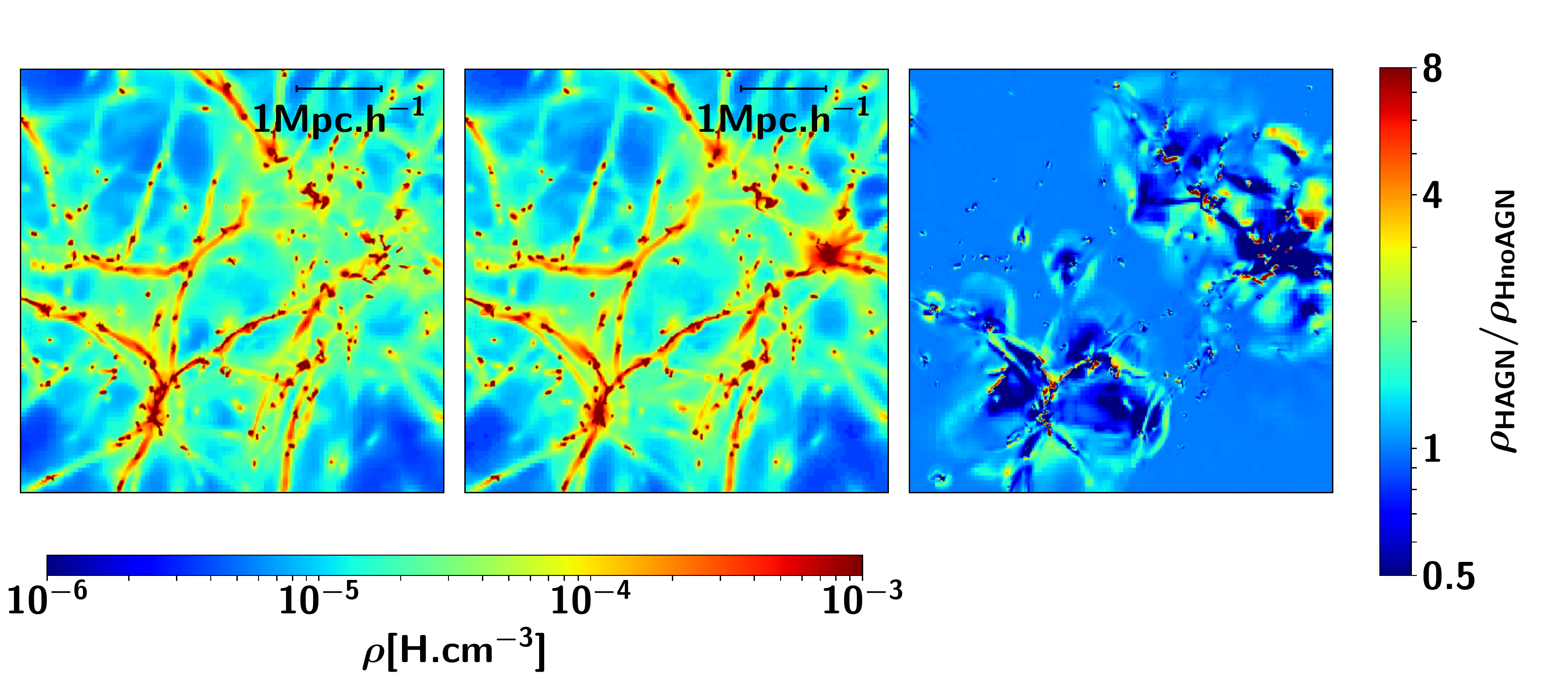}}
        \qquad
        {\includegraphics[width=0.5\textwidth,  height=3.5cm,keepaspectratio]{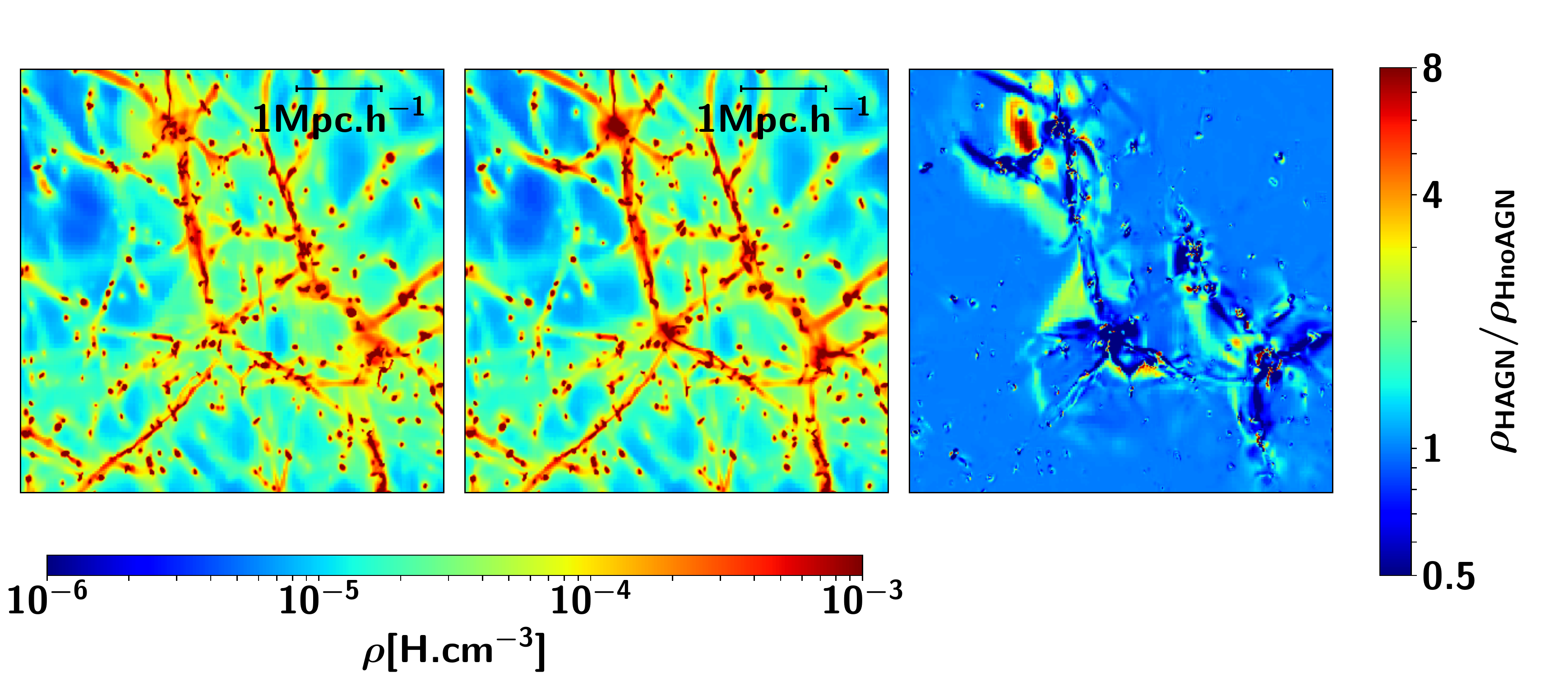}}
        {\includegraphics[width=0.5\textwidth,  height=3.5cm,keepaspectratio]{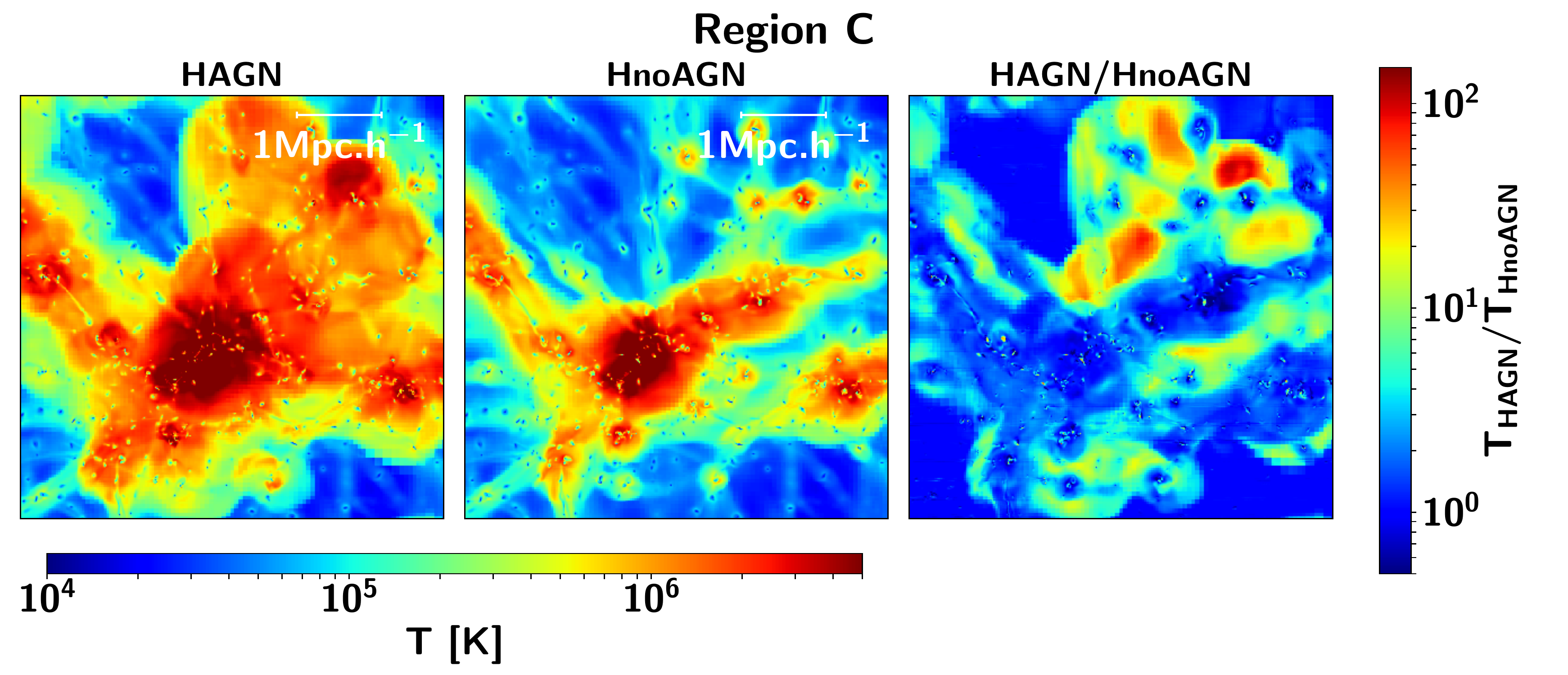}}
        \qquad
        \vspace{-1.5mm}
        {\includegraphics[width=0.5\textwidth,  height=3.5cm,keepaspectratio]{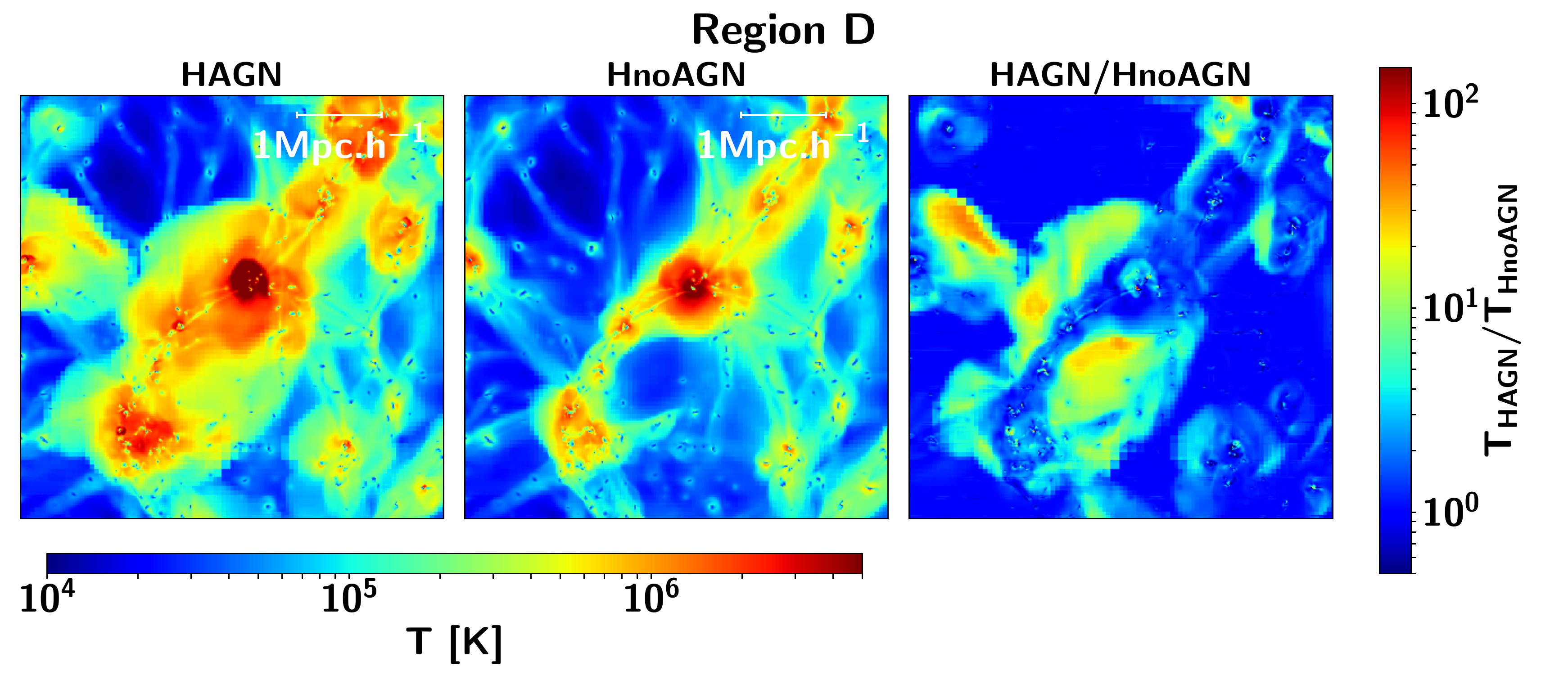}}
        {\includegraphics[width=0.5\textwidth,  height=3.5cm,keepaspectratio]{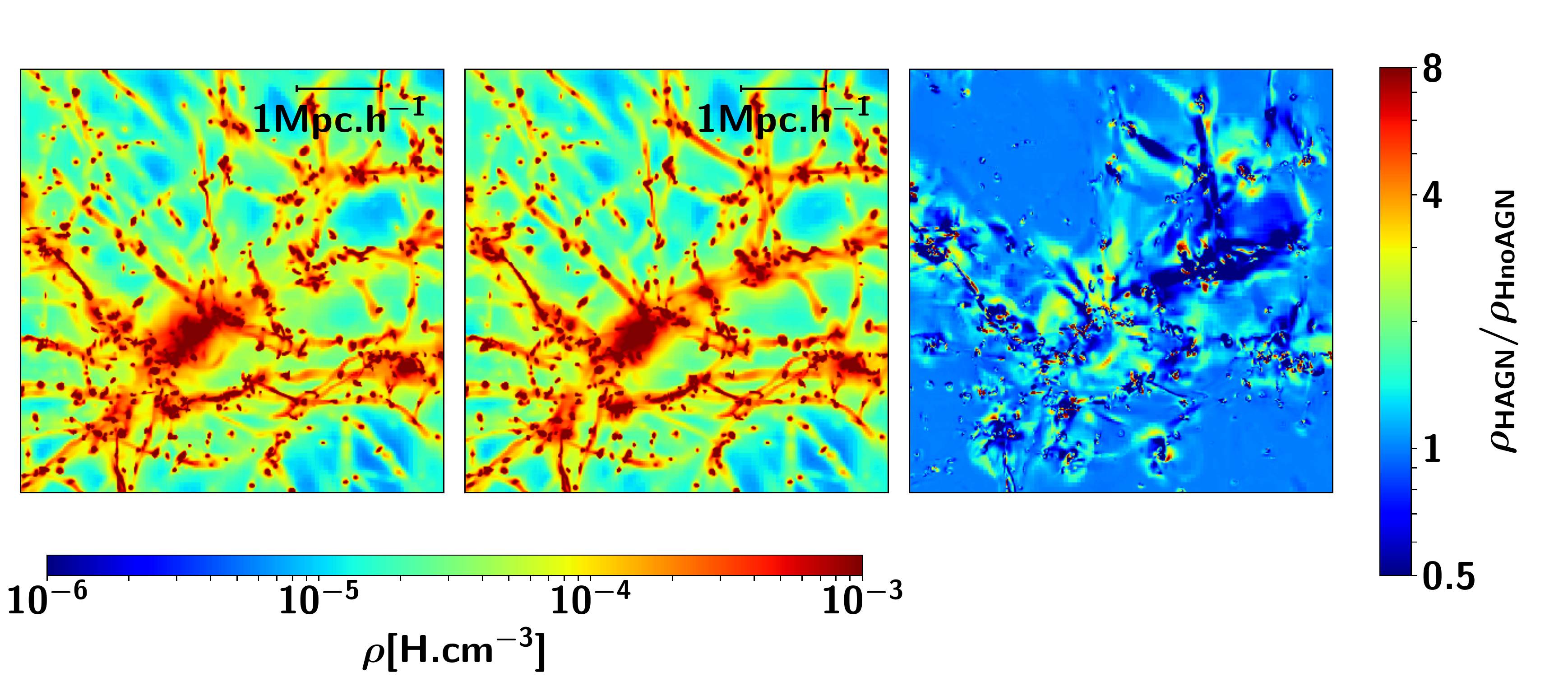}}
        \qquad
        {\includegraphics[width=0.5\textwidth,  height=3.5cm,keepaspectratio]{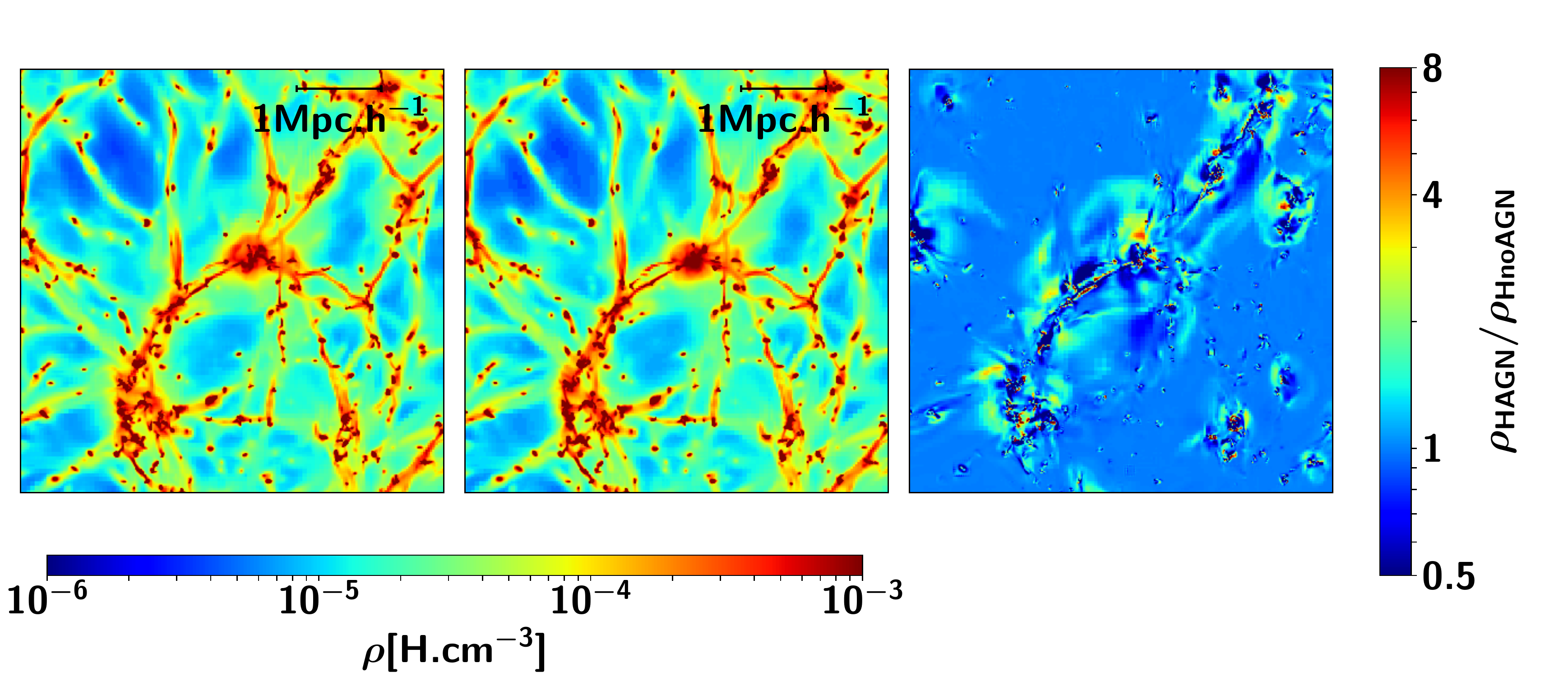}}
    \end{minipage}
    \caption{Projected temperature (first and third lines) and density (second and fourth lines) maps of \hagn \, (first and fourth columns) and \hnagn \, (second and fifth columns) and the ratio of \hagn \, over \hnagn \, for the two quantities (third and sixth columns) for the four circled regions from Fig.~\ref{fig:Trmap}. Temperature and density are encoded in $\mathrm{\log(T)}$ and $\mathrm{\log(\rho)}$ unit. Boxes are 5 $\mathrm{Mpc.h^{-1}}$ in comoving coordinate.}
    \label{fig:Tr_all}
\end{figure*}

The redistribution of gas is subtle but distinguishable on the projected density maps of \hagn \, and \hnagn \, on the bottom panels of Fig.~\ref{fig:Trmap}. Dense gas bubbles around dark matter haloes are less confined and spread to larger radius in \hagn \, compared to \hnagn. It is clearly visible on the zoom figures of Fig.~\ref{fig:Tr_all} described later.
On the temperature-density diagrams of Fig.~\ref{fig:hist2d}, the under-dense region, i.e. $\mathrm{ \log(\delta) < 2}$, is more populated in the AGN feedback case, to the detriment of the dense region, i.e. $\mathrm{\log(\delta) > 2}$. Indeed, The hot and cold IGM contain 86\% of the mass versus 82\% of the mass for \hagn \, and \hnagn \, respectively.

\begin{figure}
\begin{center}
\includegraphics[width=\columnwidth]{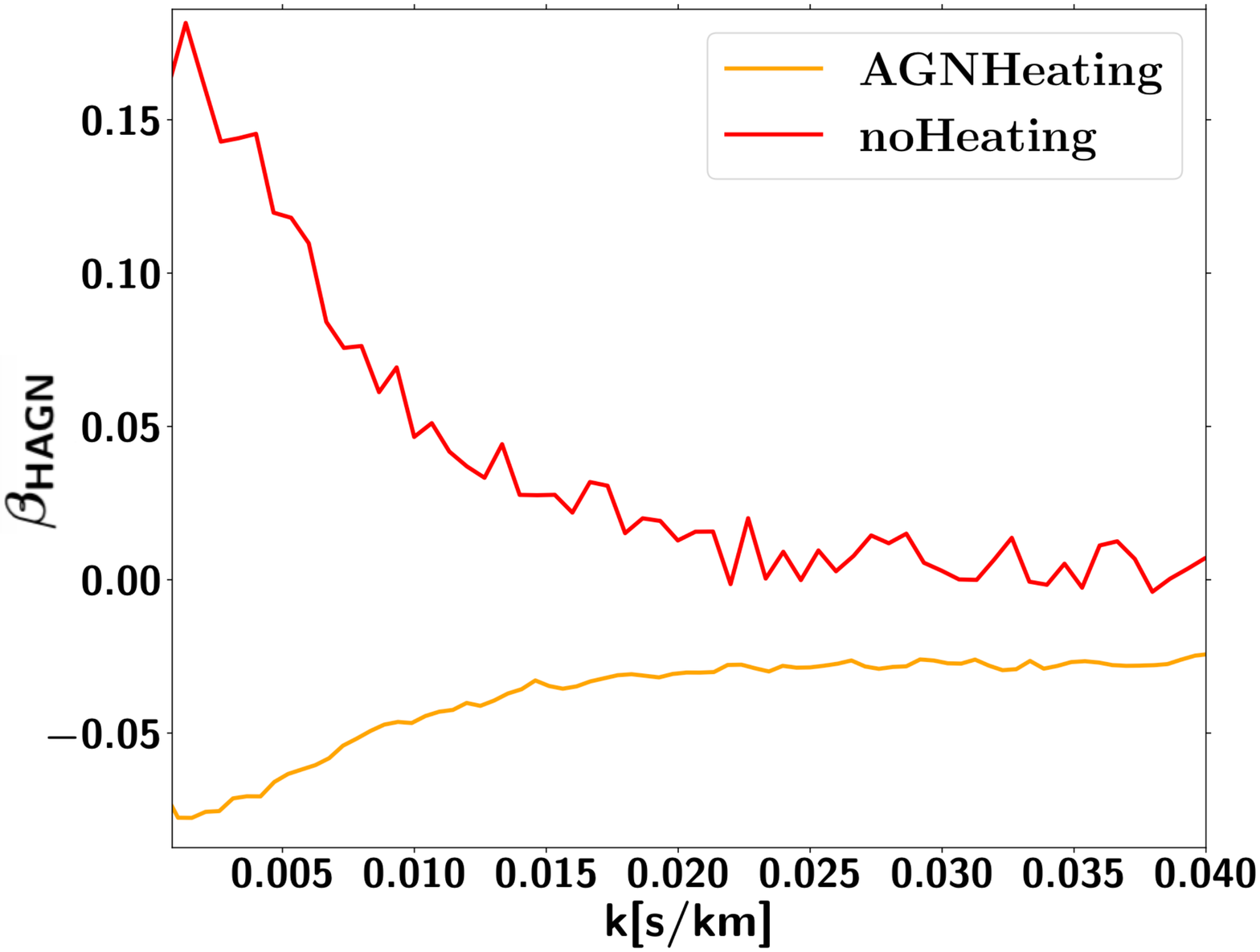}
\caption{Correction with and without heating of AGN feedback at $z=2$. The 'noHeating' curve is obtained by imposing in \hagn \, the same probability distribution function of temperature in function of density than in \hnagn \,, then we compute the neutral fraction of hydrogen.}
  \label{fig:rescaleT}
\end{center}
\end{figure}

To disentangle the gas heating and mass redistribution effects, we compute the same correction where the heating is switched-off in \hagn. To do so, we impose the \hagn \, temperature-density diagram to be the same than \hnagn. We estimate the probability distribution functions (PDF) of the temperature in 100 density bins in \hnagn. Then, for each gas particle we draw a temperature from the appropriate temperature PDF depending on the gas density. We introduce noise by decorrelating the temperature at the very small scales, therefore we also apply this modification of temperature in \hnagn. Fig~\ref{fig:rescaleT} shows this correction 'without heating' from AGN feedback. The increase of power on large scales is coherent with the study from \cite{Chisari2018}. AGN feedback redistributes gas from the small scales to the large scales, hence the matter power spectra show a suppression of power on small scales and an enhancement on the large ones. The redistribution of gas ejected from the small scales also contributes to the suppression of power, but it has an antagonist effect on the \plya \, with the strong energy injection on the large-scale modes. On Fig.~\ref{fig:Tr_all} we show the gas temperature and gas density in \hagn \, and \hnagn \, from the four circled regions from Fig.~\ref{fig:Trmap}, we also show the ratios of the density and temperature of \hagn \, over \hnagn. In all cases, we clearly see the hot and dense outflows expeled from galaxies. The outflows extend to larger scales in the \hagn \, case and the temperature on the edges are ten to hundred times higher when the density of the outflows are about two times higher. Thus, the effect of heating considerably dominates the mass redistribution on the power spectra on the large-scale modes, as seen on Fig~\ref{fig:rescaleT}. The outflows are heated to temperature high enough so that the gas stays in the ionized state, hence reducing the power on the \plya.  Also the density inside the galaxies are about then time lower in \hagn \, compared to \hnagn, confirming the depletion of gas content in galaxies and in their surrounding, hence the reduction of power in the matter power spectra on the small-scale modes.

The redshift dependence is well understood because of the increasing capacity of BH to expel gas from haloes on our redshift interval \citep{Beckmann2017}. This is combined with the displacement of energetic gas leading to the expansion of hot-gas bubbles and a net increase of the IGM temperature.
The scale dependence arises because of the sensitivity of the power spectrum modes to different regions of the baryonic gas. The large scale modes are sensitive to the diffuse gas, i.e. the efficiently heated region of the temperature-density diagram with temperature above $\mathrm{10^8 K}$, which  therefore stays hot and ionized as it is hard for hydrogen to recombine. The small-scale modes are dominated by the signal of dense regions, which are not as efficiently heated as the most diffuse regions, and can partially radiate away the injected energy and ultimately recombine. It therefore alleviates the suppression of power on the smallest scales.

 Our results are in agreement with the study done in V13 which shows a suppression of power on the large scales as well. It is also stated that it is due to the heating induced by the AGN feedback for the following reasons. First, the flux PDF  exhibits an increase of the number of pixels with high-flux transmittivity, secondly, an increase of low density gas with $\mathrm{T > 10^5 K}$ in the temperature-density diagram is observed.

\begin{figure}
\begin{center}
\includegraphics[width=\columnwidth]{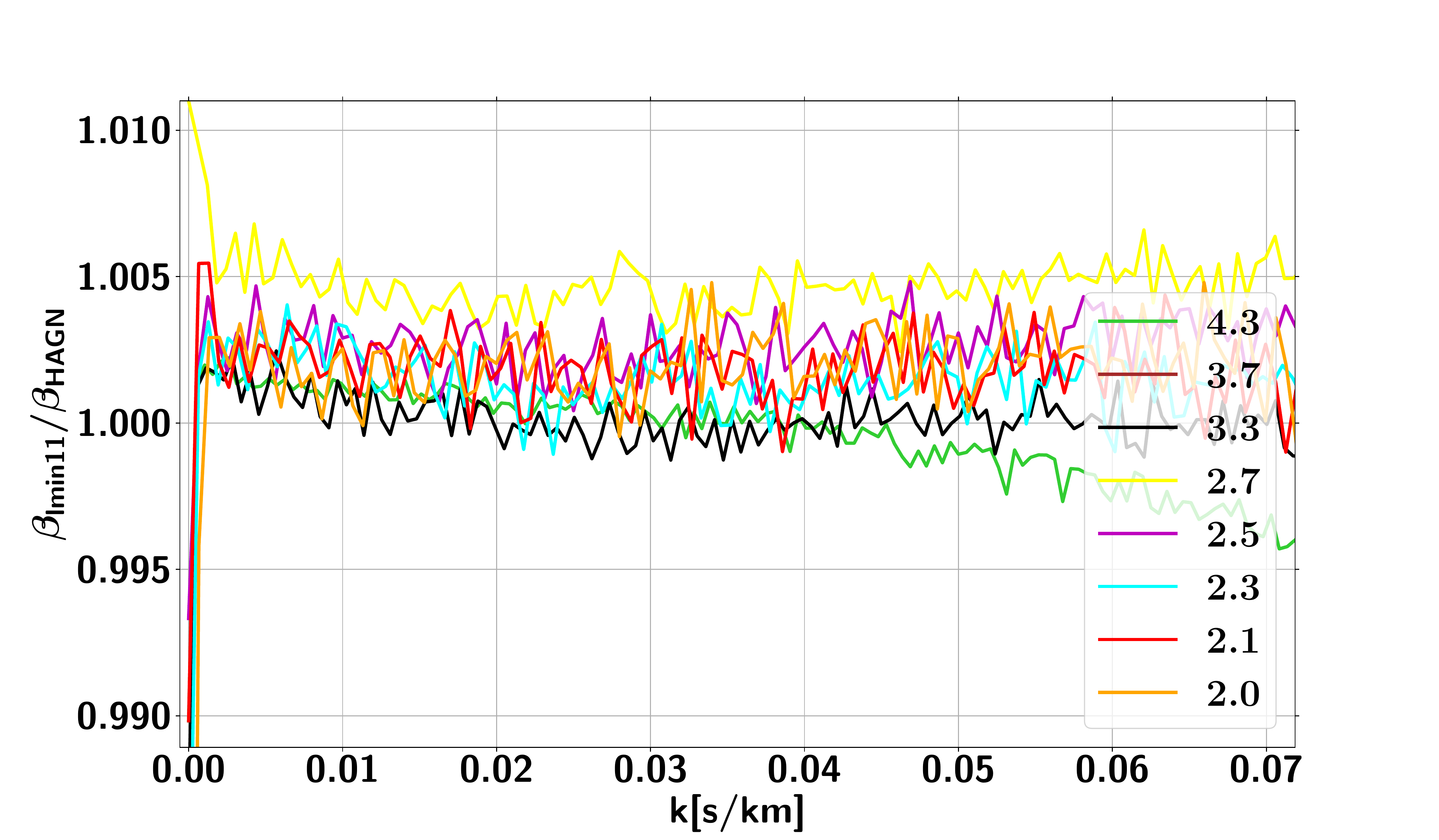}
\caption{Ratio of the more IGM-resolved simulations correction, $\beta_{\rm lmin11}$, to the fiducial correction $\beta_{\rm HAGN}$.}
  \label{fig:lmin11}
\end{center}
\end{figure}

We presented in Sec.~\ref{sec:simu} the advantages and drawbacks of \hagn \, as the fiducial simulation. If studies have been performed on the convergence of the \plya \, on uniform grids requiring 20 $\mathrm{kpc.h^{-1}}$, no convergence tests have been done, to our knowledge, when activating the AMR. Therefore we use the restarts \hlmin \, and \hnlmin \,  where we forced the refinement from level 10 to level 11 of the coarsest gas cells, to check the convergence of our corrections. Fig.~\ref{fig:lmin11} shows the differences induced by this increase of IGM resolution, with the ratio of the corrections of the \hlmin\ simulation, $\beta_{\rm lmin11}$, to the fiducial correction $\beta_{\rm HAGN}$. The resolution effect is well below the percent level at every redshifts. We stress that it does not imply that the absolute \plya \, are converged but it means that the coupling of AGN feedback and resolution is greatly subdominant when compared to other sources of uncertainties. We can therefore consider afterward that our corrections are converged on our range of scales and redshifts. The IGM resolution effects on the \lya\ forest and galaxy evolution processes is the focus of a follow-up project (Chabanier et al. in prep).

\subsection{Uncertainties}
\label{sec:uncertainties}
We identify two possible sources of uncertainties in our correction: uncertainties related to the feedback model or due to the sampling of the LOS.

Uncertainties in the feedback model are related to the uncertainties in the three main subgrid feedback parameters presented in Sec.~\ref{sec:HAGN}; the stochasticity in the accretion rate related to the boost factor $\mathrm{\alpha}$, the efficiency \ef \, and the radius of energy deposition \ragn. We use the set of additional simulations presented in Sec.~\ref{sec:variants} to estimate variations in the corrections $\beta$ at all redshifts due to fluctuations in these parameters. We arbitrary define the $\mathrm{1\sigma}$ bound due to each parameter variation as
\begin{equation}
  \sigma_i = \frac{\beta_{i}-\beta_{HAGN}}{n},
\label{eq:sig_param1}
\end{equation}
where $i$ is either $\mathrm{clp10}$, $\mathrm{clp100}$, $\mathrm{r+}$, $\mathrm{r-}$, $\mathrm{\epsilon_f+}$ or $\mathrm{\epsilon_f-}$. We take $n$ as the number of observational uncertainties between the galaxy properties measured in the simulations and the ones from observations. For \clp \, and \clpp \, we take $n = 1$ because the deviations between  properties of the two simulations and observations are at about the sigma level in terms of observational uncertainties for both the mean fraction of gas and the $\mathrm{M_{BH}}-\mathrm{M_*}$ relation. For \rp, \hrm, \ep \, and \hem \,  we take $n = 3$. Indeed,at least one of the observable is in deviation of at least $3\sigma$ in terms of observational uncertainties. We could take $n = 4$ or $n = 5$ if we combined the two probes, but because they are not fully independent and in order to be conservative we choose to keep $n =3$.

\begin{figure*}
\begin{center}
\includegraphics[width=\textwidth]{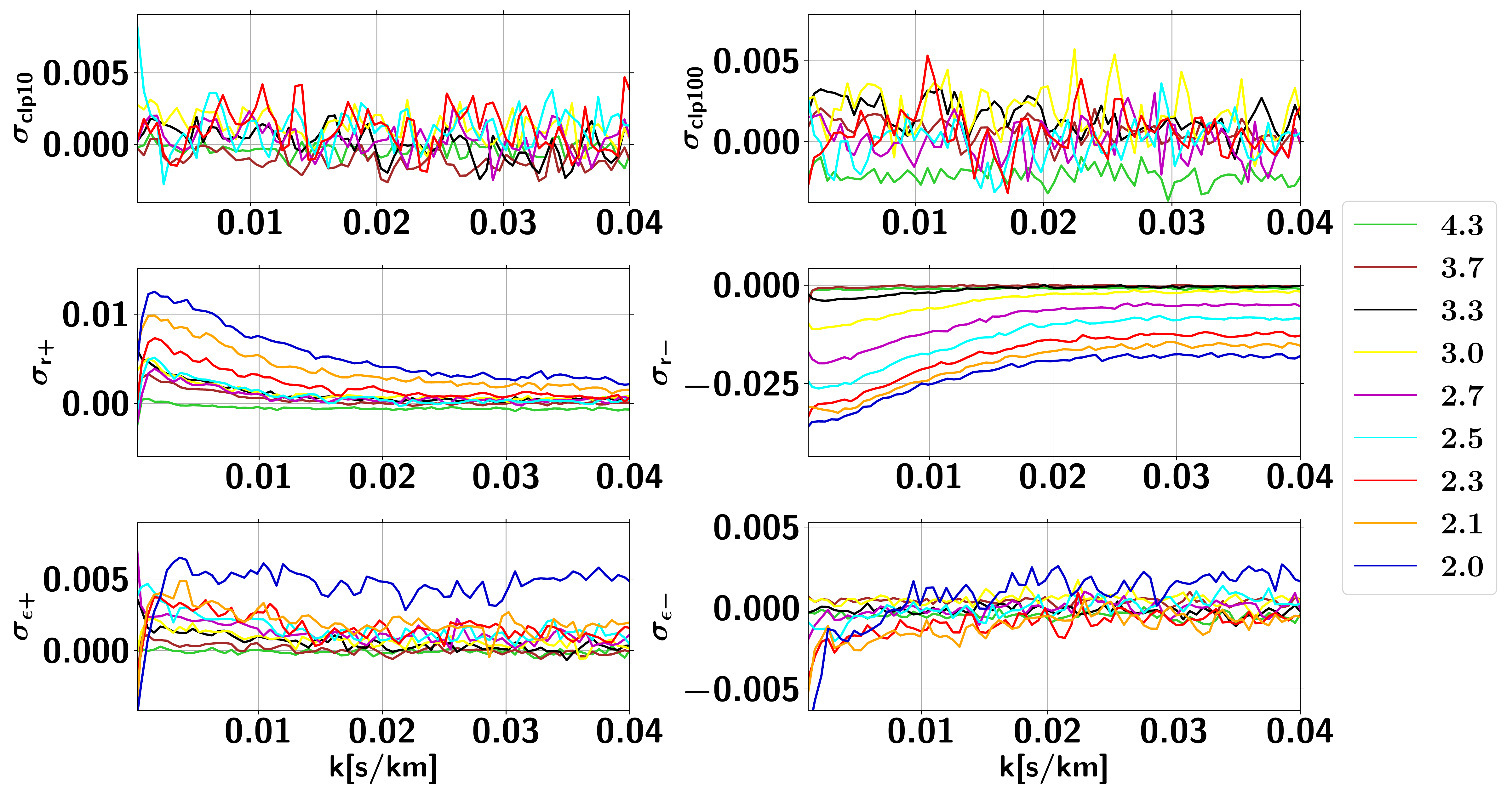}
\caption{Uncertainties $\sigma_i$ due to the three feedback parameters. From left to right and top to bottom: \clp, \clpp, \rp, \hrm, \ep and \hem, the different colors are for the nine redshifts.}
  \label{fig:sigs}
\end{center}
\end{figure*}

\begin{figure*}
\begin{center}
\includegraphics[width=\textwidth]{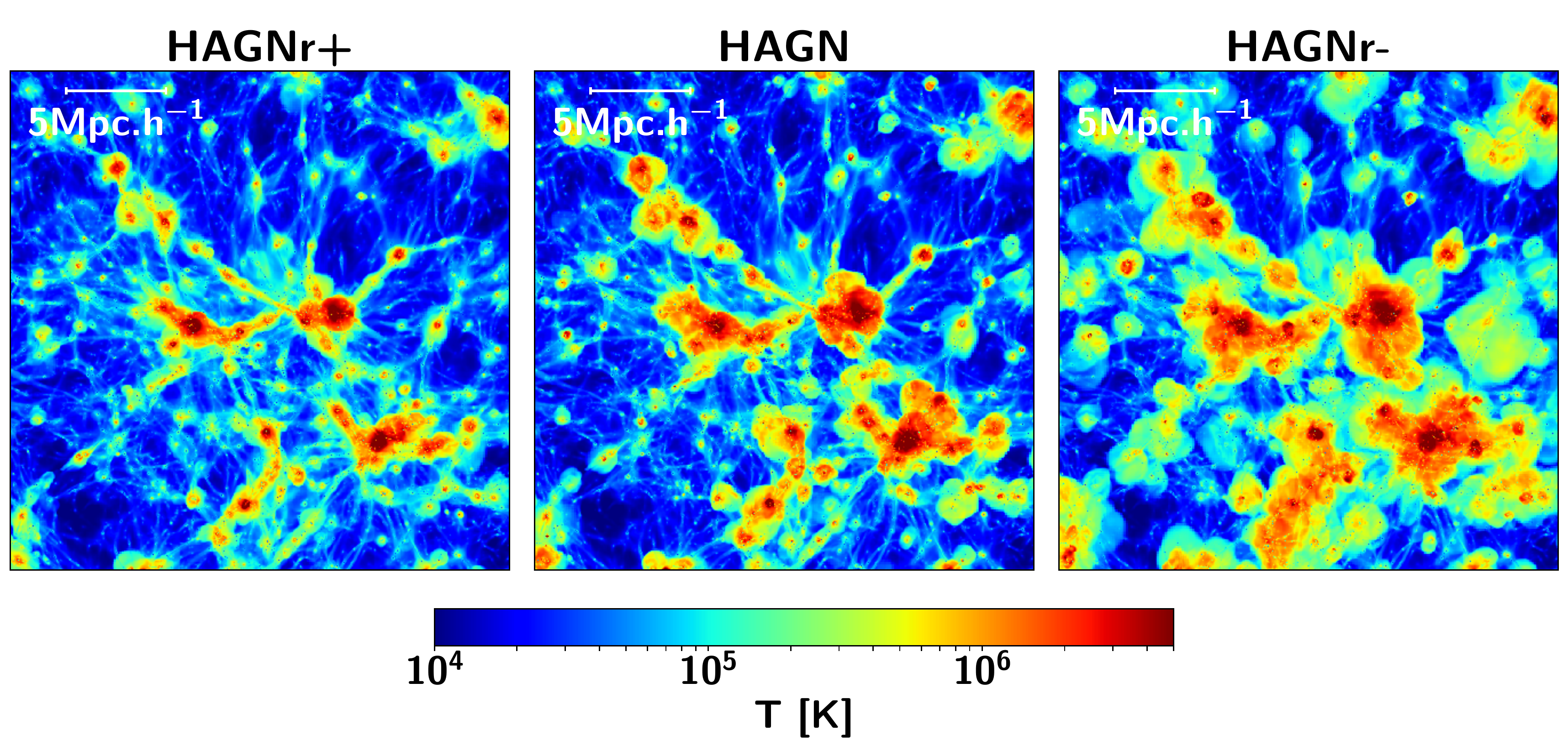}
\caption{Projected temperature maps of \rp, \hagn \, and \hrm \, from left to right at $z = 2.0$ encoded in $\mathrm{\log(T)}$ unit. Boxes are 25 $\mathrm{Mpc.h^{-1}}$ in comoving coordinate.}
  \label{fig:T_ragn}
\end{center}
\end{figure*}

 Fig.~\ref{fig:sigs} shows the $1\sigma$ bound of each of the parameters, at every redshifts:
\begin{itemize}
  \item \clp \, and \clpp \, results are presented on the left and right panels of the first row. The stochasticity introduced in the accretion rates of BH appears not to have any noticeable effect on the \plya \, as the deviations are well below the percent level. Therefore we do not consider any uncertainty due to $\mathrm{\alpha}$ in the following.
  \item \rp \, and \hrm \, are on the second row. The parameter \ragn \, comes out to be the one to which \plya \, is the most sensitive to, with deviations up to 1\% for the upper bound and up to 4\% for the lower bound. We show that lowering \ragn \, leads to a stronger feedback than increasing it. Indeed, \hrm \, shows a significant decrease of power compared to \rp \, which displays an increase of power on large scales. Giving more energy to a smaller volume and keeping the same amount of injected energy produces larger hot bubbles of gas around AGNs as we illustrate in the temperature maps of \rp \, and \hrm \, at $z=2$ in Fig.~\ref{fig:T_ragn}. There is therefore more ionized gas on large scales when \ragn \, is lower. This result is in opposition with \cite{Dubois2012} that shows less ionization for low \ragn. We put this on the account of a different feedback prescription; in \cite{Dubois2012} the energy injection was volume weighted when it is mass weighted in \hagn. In the first case, when we broaden the region of energy deposition, we impact more diffuse cells that are equally heated than the dense ones and are less likely to radiate away the injected energy, leading to an increase of the ionized region. However in the later case, diffuse cells get less energy than the dense ones, then the dilution makes the feedback less effective.
  \item \ep \, and \hem \, are on the third row. Modifications in \ef \, do not impact the flux power spectrum above the percent level. Even if the $\mathrm{M_{BH}}-\mathrm{M_*}$ and $\mathrm{ f_{gas}}$ of the two resimulations largely differ from observations, the self-regulation of the BH prevents large modifications of the IGM thermal state. We can identify a trend, but since it is largely sub-dominant compared to \ragn \, we do not consider any effects due to \ef \, afterward.
\end{itemize}

To estimate uncertainties due to the sampling of our LOS sample we compute the root-mean-square (RMS) error of the corrections from five different sets constitued of 20.$\mathrm{10^3}$ LOS. For $z = 4.25$ and $z = 2$ it leads to uncertainties  at the level of $\mathrm{10^{-3}}$. This is subdominant compared to the uncertainties due to \ragn. Therefore we do not consider statistical uncertainties in the following. \newline

\begin{figure*}
\begin{center}
\includegraphics[width=\textwidth]{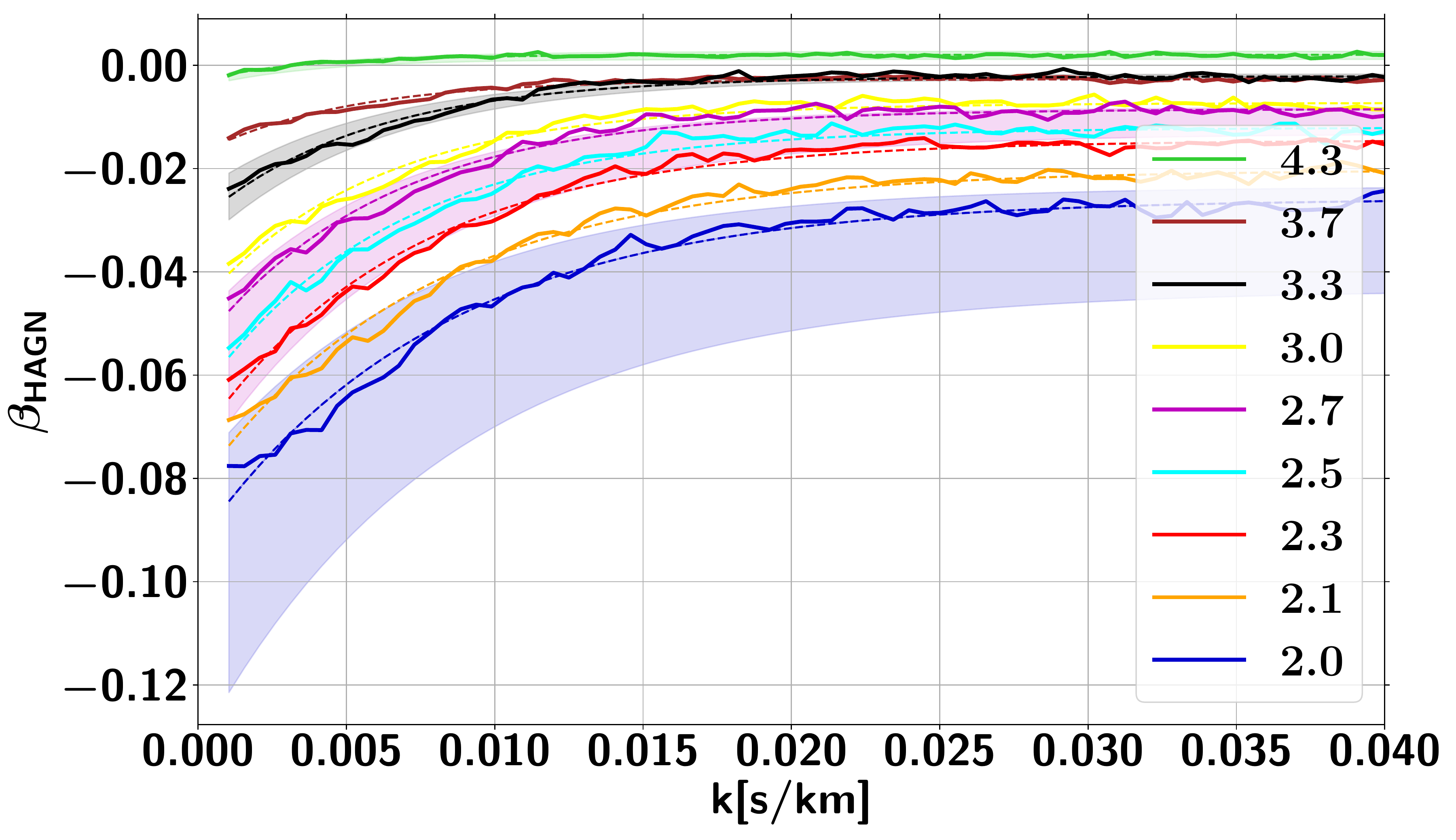}
\caption{Corrections at all redshits and the associated fits in dashed lines. Uncertainties are displayed in the shaded areas at $\mathrm{z=4.25}$, $\mathrm{z3.3}$, $\mathrm{z=2.7}$ and $\mathrm{z=2.0}$ only for the sake of readibility. Ucertainties are systematics related to uncertainty in the feedback model only.}
  \label{fig:corrfits}
\end{center}
\end{figure*}

We showed that the impact of AGN feedback on the flux power spectra is to globally suppress the power at all scales. The suppression is explained by the combination of an efficient heating and by the mass redistribution from small to large scales. The suppression is enhanced with decreasing redshifts because of the increasing capacity of BH to expel gas from haloes and the displacement of hot gas, which induce a stronger feedback. The scale dependence arises because  the large-scale modes are sensitive to the diffuse gas and the small-scale modes are dominated by the signal of dense gas that can partly radiate away the injected energy,  which alleviates the suppression.
The uncertainty on our correction is strongly dominated by the radius of energy deposition \ragn, because stochasticity in the accretion does not appear to be efficient in our redshift range, and variation of the efficiency is counter-balanced by self-regulation of the BH. It demonstrates that the efficiency of the heating, hence the ionization, has more effect than the amount of injected energy in the medium. We provide analytical fits for the corrections and upper and lower bounds defined as $\mathrm{\beta(HAGN)^{+\sigma_{r+}} _{-\sigma_{r-}}}$. We fit our corrections with the following function, $ f(k) = a + b \exp(-ck)$, where the parameters $a$, $b$ and $c$ are given for the three fits at all redshifts in Tab.~\ref{tab:corr_fits}. We also provide this table online as an ascii file in the accompanying material attached to the paper. We show in Fig.~\ref{fig:corrfits} the fits of the correction in dashed lines, we also display the upper and lower uncertainties in the shaded areas at $\mathrm{z=4.25}$, $\mathrm{z=3.3}$, $\mathrm{z=2.7}$ and $\mathrm{z=2.0}$ only for the sake of readibility.

\begin{table*}
\centering
\begin{tabular}{c||ccc|ccc|ccc}
\hline
Redshift & $a$ & $b$ & $c$ & $a_+$ & $b_+$ & $c_+$ & $a_-$ & $b_-$ & $c_-$ \\
\hline
4.25 & 2.03\et & -5.03\et & 2.74\edd & 2.71\et & -5.23\et & 2.26\edd & 1.16\et & -5.49\et & 2.71\edd \\
3.7 & -2.70\et & -1.46\ed & 2.05\edd & -2.60\et & -1.07\ed & 2.06\edd & -2.98\et & -1.54\ed & 2.30\edd \\
3.3 & -2.13\et & -2.81\ed & 1.79\edd & -1.68\et & -2.31\ed & 1.77\edd & -2.51\et & 3.27\ed & 1.70\edd \\
3.0 & -7.29\et & -3.94\ed & 1.70\edd & -6.66\et & -3.44\ed & 1.64\edd & -8.86\et & -5.09\ed & 1.51\edd \\
2.7 & -8.46\et & -4.63\ed & 1.61\edd & -8.11\et & -4.20\ed & 1.58\edd & -1.32\ed & -6.47\ed & 1.40\edd  \\
2.5 & -1.21\ed & -5.23\ed & 1.56\edd & -1.19\ed & -4.65\ed & 1.55\edd & -2.00\ed & -7.44\ed & 1.34\edd \\
2.3 & -1.45\ed & -5.81\ed & 1.43\edd & -1.38\ed & -5.03\ed & 1.45\edd & -2.64\ed & -8.07\ed & 1.27\edd \\
2.1 & -2.02\ed & -6.12\ed & 1.30\edd & -1.85\ed & -5.14\ed & 1.35\edd & -3.48\ed & -8.32\ed & 1.19\edd \\
2.0 & -2.58\ed & -6.66\ed & 1.23\edd & -2.35\ed & -5.48\ed & 1.33\edd & -4.35\ed & -8.84\ed & 1.20\edd \\
\hline

\end{tabular}
\caption{Parameters of the correction fits on the function $f(k) = a + b \exp(-ck)$ with one line per redshift. The parameters $a$, $b$ and $c$ are for corrections given by \hagn, $a_+$, $b_+$ and $c_+$ are for the upper bound fits, and $a_-$, $b_-$ and $c_-$ are for the lower bound fits.}
\label{tab:corr_fits}
\end{table*}

\subsection{Impact on cosmological parameters}
\label{sec:cosmo_impact}
\input{cosmo}

%% file: cosmo.tex
To highlight the impact of AGN feedback on \plya \, when performing cosmological parameter inferences, we compute cosmological parameters in the four following situations: without applying AGN correction or applying the three corrections that span our uncertainty interval, i.e. the upper (weakAGN), fiducial (fidAGN) and lower (strongAGN) corrections.

We use data from the eBOSS DR14 release~\citep{DR14eB}, corresponding to the entirety of the BOSS survey complemented by the first year of eBOSS. We take the 1D transmitted flux power spectrum measured by the BOSS and eBOSS collaborations in \cite{Chabanier2019a} corrected for SN feedback with the correction from V13. The theoretical predictions of \plya \, come from a set of hydrodynamical simulations described in \cite{Borde2014}, run using a  parallel tree smoothed particle hydrodynamics  (tree-SPH) code {\sc Gadget-3}, an updated version of the public code  {\sc Gadget-2} \citep{Springel2005}. The simulations were started at $z=30$, with initial transfer functions and power spectra computed with {\sc CAMB}~\citep{Lewis2000}, and initial particle displacements generated with second-order Lagrangian Perturbation Theory.  Two particle types were included: collisionless dark matter, and gas.

We use the same likelihood than in \cite{Chabanier2019a}. It is built upon three categories of parameters. The cosmological parameters are based on a $\mathrm{\Lambda}$CDM cosmology with $H_0$, $\Omega_m$, $n_s$ and $\sigma_8$. The astrophysical parameters are chosen to follow the evolution of the IGM thermal state. The temperature-density relation is modeled by a power law with $T = T_0 \delta ^{\gamma -1}$. The evolution with redshift is modeled by a broken power-law for $T_0$, defined with $\eta^{T_0} (z<3) $ and $\eta^{T_0} (z>3)$ the two logarithmic slopes , and a simple power law for $\gamma$ with $\eta^\gamma$ the logarithmic slope. The photo-ionization rate of each simulation was fixed at each redshift to follow the empirical law $\tau_{\mathrm{eff}}(z) = A^{\tau}(1+z)^{\eta^{\tau}}$. Therefore we consider the parameters $T_0$ (z = 3), $\eta^{T_0} (z<3)$, $\eta^{T_0} (z>3)$, $\gamma$, $\eta^\gamma$, $A^{\tau}$, $\eta^{\tau}$ and two amplitudes for the correlated absorption between Ly$\alpha$ and $\mathrm{Si^{+}}$and $\mathrm{Si^{2+}}$. Finally, the nuisance parameters allow to account for uncertainties in the spectrograph resolution, residuals of Damped Ly$\alpha$ absorbers or BALs, bias from the splicing technique or UV fluctuations. We do not introduce any nuisance parameters for the AGN feedback because we directly correct our theoretical predictions with the functions given in Tab.~\ref{tab:corr_fits}. We stress that the common practice is to introduce additional parameters to be fitted along with cosmological parameters and to be marginalized over. However, this study aims at being illustrative to show the impact of the fiducial correction on the first hand, and, on the other hand, the variation of the cosmological prameters on our uncertainty range.

\begin{figure}
\begin{center}
\includegraphics[width=\columnwidth]{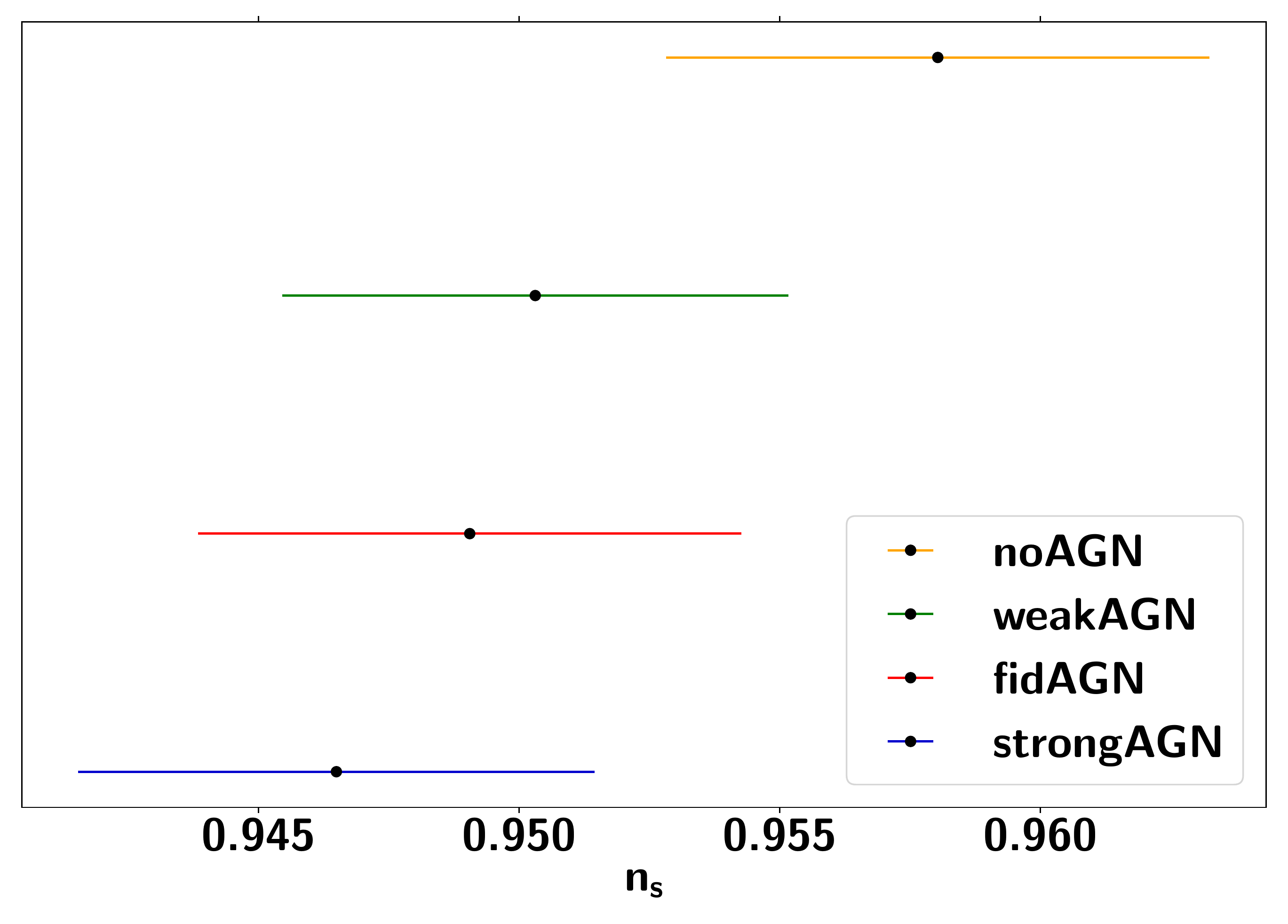}
\caption{Scalar spectral index bias induced by baryonic feedback on the 1D power spectrum from the Ly$\alpha$ forest. We show the inferred values for the four following cases: no AGN correction, the weakAGN, fidAGN and strongAGN corrections.}
  \label{fig:ns}
\end{center}
\end{figure}

Tab.~\ref{tab:cosmo} presents the best-fit values for the four cases: without AGN correction, with the weakAGN, fidAGN or strongAGN corrections. The most impacted cosmological parameter is the scalar spectral index $n_s$. We show in Fig.~\ref{fig:ns} the inferred values $n_s$ for the four configurations. It is an expected result as AGN feedback tends to increase the slope of the flux power spectrum, it is therefore degenerate with $n_s$. Not taking into account AGN feedback yields a bias of about 1\% which represents two standard deviations $\mathrm{\sigma_{stat}}$. However, it varies on less than 0.5\% on our uncertainty range in spite of the large suppression range allowed at low redshift. The bias induced by AGN feedback should have repercussion on the sum of the neutrinos masses, as a large correlation of 50\% exists between the two cosmological parameters \citep{PalanqueDelabrouille2015b}. The bias reaches 2\% for $\sigma_8$, but because it is less constrained, the shift is contained within the statistical error. Finally $\Omega_m$ does not present any significant deviation. This is also expected because, as shown in Fig.9 of~\cite{Borde2014}, varying $\Omega_m$ impact the formation of small-scale structures, hence it has more significant impact at large k's.
The astrophysical parameters are more impacted with biases of 1\% and from 3\% to 6\% for $T_0$ and $\gamma$ respectively. However it stays contained in the uncertainty range as the IGM thermal state is not well constrained by the medium-resolution eBOSS data. Also, as already stated, the AGN feedback increases the mean flux  with a shift from 2\% to 3\% on the effective optical depth amplitude which represents $1\mathrm{\sigma_{stat}}$. Finally the uncertainties on cosmological and astrophysical parameters are hardly impacted by the AGN feedback. The uncertainty on the amplitude temperature at $z = 3$ decreases from 2.0 to 1.2 when applying the strongAGN correction. However, the thermal history is described by a total of five parameters showing significant correlations, e.g. 75\% for $T_0(z=3)$ and $\gamma (z=3)$. While the uncertainty on $T_0(z=3)$ is decreased, the uncertainties on $\gamma (z=3)$ and on the redshift evolutions are all increased, which mitigates the conclusion on the precision of the estimated temperature.

\begin{table*}

\centering
\begin{tabular}{l|cccc}
  \hline
  & noAGN & weakAGN & fidAGN & strongAGN \\
\hline
$\sigma_8$ \dotfill & $0.82 \pm 0.02$ & $0.83 \pm 0.02$ & $0.83 \pm 0.02$ &  $0.83 \pm 0.02$\\
$n_s$ \dotfill& $0.958 \pm 0.005$ & $0.950 \pm 0.005$ & $0.949 \pm 0.005$ & $0.946 \pm 0.005$\\
$\Omega_m$ \dotfill& $0.268 \pm 0.009$ & $0.269 \pm 0.009$ & $0.270 \pm 0.009$ & $0.269 \pm 0.009$ \\
\hline
$T_0(z=3)$ ($10^3$ K) \dotfill & $8.5 \pm 2.0$& $8.6 \pm 1.8$ & $8.64 \pm 1.9$  & $8.7 \pm 1.2$ \\
$\gamma$ \dotfill& $0.92 \pm 0.13 $ & $0.95 \pm 0.12$ & $0.93 \pm 0.14$ & $ 0.97 \pm 0.15$\\
$A^{\tau}$ ($10^{-3}$) \dotfill& $2.33 \pm 0.06$ & $2.37 \pm 0.06$ & $2.38 \pm 0.06$ & $2.40 \pm 0.06$\\
$\eta^{\tau}$ \dotfill& $3.83 \pm 0.03$ & $3.83 \pm 0.03$ & $3.84 \pm 0.03$ & $3.84 \pm 0.03$ \\
\hline

\end{tabular}
\caption{Best-fit values and $68 \%$ confidence levels of the cosmological and astrophysical parameters of the model fitted to \plya \, measured with the SDSS data. In the first column no correction for the AGN feedback is applied, we apply the upper bound (weakAGN), the fiducial (fidAGN) and the lower bound (strongAGN) correction in the second, third and fourth columns respectively.}
 \label{tab:cosmo}
\end{table*}

%% file: discussion.tex
In this paper we present the signatures of AGN feedback on the 1D power spectrum of the \lya \, forest given that the 1\% precision reached in the measurements requires an improvement of theoretical predictions from hydrodynamical simulations. To do so,  we use the cosmological hydrodynamical Horizon-AGN simulation \citep{Dubois2014}, run with the AMR code RAMSES in a $\mathrm{L_{box} = 100 Mpc.h^{-1}}$ box. Along with \hagn \, and its no-AGN version, we performed a suite of 6 additional simulations that cover the whole plausible range of feedback and feeding parameters according to the resulting galaxy properties. We choose \hagn \, as the fiducial simulation because it presents the adequate caracteristics in termes of box size and resolution to accurately reproduce the \lya \, forest, and  because it is in fairly good agreement with some key observational galaxy properties, which is necessary to explore realistic feedback models. The series of 6 additional simulations modify either the boost factor $\alpha$ and its stochasticity, the radius of energy deposition \ragn \, or the fraction of radiated energy injected in the medium \ef. The set of feedback and feeding parameters was chosen to span observational uncertainties of the  $\mathrm{M_{BH}}-\mathrm{M_*}$ relation and of the mean fraction of gas in galaxies at $z=2$.

We show that the impact of AGN feedback on the P1D is to globally decrease the power at all scales. This effect increases with decreasing redshift. The suppression is explained by the combination of an efficient heating and by the mass redistribution from small to large scales. The suppression is enhanced with decreasing redshifts because of the increasing capacity of BH to expel gas from haloes and the displacement of hot gas. The scale dependence arises because the large-scale modes are sensitive to the diffuse gas and the small-scale modes are dominated by the signal of dense gas that can partly radiate away the energy injected by AGNs, thus  alleviating the suppression. The uncertainty is strongly dominated by \ragn. Introducing stochasticity in the accretion rate with $\alpha$ appears inefficient within our redshift range. And because the BHs self-regulate its mass and accretion rate, injecting more or less energy in the medium, i.e. modifying \ef, do not impact the \lya \, forest. It demonstrates that the efficiency of the heating, hence the ionization, has more effects than the amount of injected energy in the medium. Our results are consistent with the study done in V13 showing that the redistribution of mass and energy induces a suppression of power on the large scales of the \plya.

We provide analytical fits for the corrections and upper and lower bounds of the uncertainties. We fit our corrections with a function $ f(k) = a + b \exp(-ck)$, where the parameters $a$, $b$ and $c$ are given for the three fits at all redshifts in Table.~\ref{tab:corr_fits}. We also provide this table online as an ascii file in the accompanying material attached to the paper.
Using these corrections make it possible to account for AGN effects at post-processing stage in future work given that running simulations without AGN feedback can save considerable amounts of computing resources.

We test the impact of our corrections and show that ignoring the effects of AGN feedback in cosmological analysis using the \plya \, leads to 1\% biases on the scalar spectral index $n_s$ which represents two times the current statistical uncertainty on this parameter. The bias reaches 2\% for $\sigma_8$, but because it is less constrained, the shift is contained within the statistical error. However, in spite of the large uncertainty of our AGN correction the biases vary on less than 0.5\% within our uncertainty range. The biases are more significant for the astrophysical parameters related to the temperature of the IGM and the optical depth with shifts up to 6\%. However it is well contained in the statistical uncertainties because the IGM thermal state is not well constrained by the eBOSS medium-resolution data used in this study.

We have shown that the effects of AGN feedback are not negligible given the level of precision reached by the data. However this is not the only baryonic process impacting the \lya \, forest; SN feedback, gas cooling and structuration in the IGM also need to be taken into account to properly model the \plya \, in hydrodynamical simulations considering that uncertainty errors will even shrink further with the advent of spectroscopic surveys such as DESI. We leave for future works to properly estimate the impact of SN feedback, but also the intertwining of stellar and AGN feedback, which has been shown to be significant on the galaxy properties, such as the SFR density \citep{Biernacki2018}, and finally any cosmological dependence of such mechanisms.